\magnification1200

\tolerance=1200

\baselineskip14pt

\font\small=cmr10 at 10truept

\font\reg=cmr10 at 12truept

\font\large= cmr18 at 18truept

\font\llarge=cmr20 at 21truept

\font\lllarge=cmr24 at 24truept

\rightline{KCL-MTH-06-13} \rightline{hep-th/0611318}

\vskip 2cm

\lllarge

\centerline{  Duality Groups, Automorphic Forms}

\centerline{ and Higher Derivative Corrections }

\vskip 2cm

\llarge \centerline{Neil Lambert  and Peter West}

\bigskip
\large \centerline{Department of Mathematics}

\centerline{King's College, London}

\centerline{ WC2R 2LS, UK}

\bigskip
\reg\centerline{neil.lambert@kcl.ac.uk\ , peter.west@kcl.ac.uk}

\bigskip

\large \centerline{\sl Abstract}

\reg We study the higher derivative corrections that occur in type
II superstring theories in ten dimensions or less. Assuming
invariance under a discrete duality group $G({\bf Z})$ we show that
the generic functions of the scalar fields that occur can be
identified with automorphic forms. We then give a systematic method
to construct automorphic forms from a given group $G({\bf Z})$
together with a chosen subgroup $H$ and a linear representation of
$G({\bf Z})$. This construction is based on the theory of non-linear
realizations and we find that the automorphic forms contain the
weights of $G$. We also carry out the dimensional reduction of the
generic higher derivative corrections of  the IIB theory to three
dimensions and find that the weights of $E_8$ occur generalizing
previous results of the authors on M-theory. Since the automorphic
forms of this theory contain the weights of $E_8$ we can interpret
the occurrence of weights in the dimensional reduction as evidence
for an underlying U-duality symmetry.

\bigskip
\reg \noindent

\vfill \eject

{\large {1. Introduction}}
\bigskip
By virtue of the large amount of supersymmetry they possess, the IIA
supergravity [1,2,3] and IIB supergravity [4,5,6] theories encode
all the perturbative and non-perturbative low energy effects of the
corresponding string theories. Furthermore, the eleven dimensional
supergravity theory [7] is thought to be the low energy effective
action for an as yet undefined theory called M-theory. The IIB
theory possess an $SL(2,{\bf R})$ symmetry [4] while the IIA
supergravity and IIB supergravity dimensionally reduced on an
$(n-1)$-torus, or equivalently the eleven-dimensional supergravity
theory on an $n$-torus, possess an $E_n$ symmetry for $n=7,8,9$
[8,9,10] and possibly $n=10$ [11].  For other work on  symmetries
that appear in dimensional reduction see [12-18] and we note that
the $E_n$ symmetries necessarily contain $T$-duality which is a
perturbative symmetry of string theory [19,20] . These theories
possess charged states which are rotated by these symmetries and
their charges obey the quantization condition [21]. This has lead to
the conjecture [22,23,24] that a discrete version of these groups,
denoted by $G({\bf Z})$, are symmetries in string theory, {\it e.g.}
$SL(2,{\bf Z})$ in the case of ten dimensional IIB string theory.

However, much of the considerations of these discrete symmetries has
been within the context of the lowest order effective action, {\it
i.e.} the  maximal supergravity theories, and  there has not been
much discussion of the role of these symmetries in the higher
derivative corrections (however see [25,26]). A particularly notable
exception to this are the higher derivative terms of the from
$D^{2k} R^4$ that occur in IIB string theory whose coefficients for
$k \leq 3$ have been determined exactly [27-33]. These coefficients
are functions of $\tau=\chi+ie^{-\phi}$, where $\chi$ is the axion
and $\phi$ the dilaton. Under the action of $SL(2,{\bf Z})$, $\tau$
is acted on by a fractional linear transformation, however  the
Riemann tensor is inert (in Einstein frame). Imposing that
$SL(2,{\bf Z})$ is a symmetry one immediately sees that these
coefficients must be invariant under $SL(2,{\bf Z})$ and hence are
given by automorphic forms. The work of [27-33] has identified the
automorphic forms for $k\leq 3$ and shown that some of the infinite
series of terms are consistent with certain explicit string theory
calculations  and, perhaps more remarkably, loop calculations in
eleven-dimensional supergravity. Furthermore the coefficients of the
$R^4$ term that occurs upon compactification on a torus to eight and
fewer dimensions have been obtained as automorphic forms of
$SL(3,{\bf R}),SL(5,{\bf R})$ and $E_n$ [34,35,36,37]. In a similar
spirit the coefficients  in eight dimensions of $R^4G_3^{4g-4}$
terms, where $G_3$ is the modified complexified three-form field
strength of type IIB string theory, have been given as automorphic
forms of $SL(3,{\bf R})$ [38].

In a recent paper [39], we explored the dimensional reduction to
three dimensions of generic higher derivative terms that arise in
eleven-dimensional M-theory. The main purpose of this work was  to
see if there are traces of the $E_8$ symmetry that is present in the
low energy effective action in three dimensions {\it i.e.} the
$N=16$ maximal supergravity theory in three spacetime dimensions.
Three dimensions is special because it is the first dimension in
which all dynamical fields are scalars (after a dualizing any
vectors modes) and in the low energy effective action these scalar
fields can be identified as a non-linear realization of $E_8$ with
local subgroup $SO(16)$. In reference [39] we determined the
dependence of arbitrary higher derivatives terms on the diagonal
components of the metric associated with the torus, which we may
parameterize by $g_{ii}=e^{ -c_i\phi_i}$ for some constant $c_i$.
These occur in the action in the form of factors $e^{\sqrt{2}\vec
v\cdot \vec \phi}$ that multiply the derivatives of the scalar
fields.  The different possible vectors $\vec v$ arise as the
different possible terms the exponential factor  can multiply. For
the lowest order effective theory the vectors $\vec v$ are just the
positive roots of $E_8$. This is readily understood from the well
known fact that the effective action can be written in terms the
Cartan form of the coset $E_8/SO(16)$ which lives in the adjoint
representation of $E_8$ and can be explicitly shown to involve the
positive roots. However, in [39] it was found that the dimensional
reduction of the higher derivative terms does not lead to the
positive roots of $E_8$. Rather one finds that the various vectors
that arise are elements of the weight lattice of $E_8$. Moreover one
only finds weights for the types of higher derivative terms that are
expected to arise in M-theory [40-57]. Weights have also appeared in
the higher derivative effective action in [58,59] within the context
of $E_{10}$  and  `cosmological billiards' [60,61].

While the occurance  of weights of $E_8$ in the dimensional
reduction of the higher derivative terms  indicates the presence of
some $E_8$ structure it was unclear what this structure could be
since the non-linear realizations to which the scalars belong are
usually constructed from the Cartan forms and these only contain the
positive roots of $E_8$.

In this paper we will show that if one assumes that the higher
derivative terms of a type II string theory in ten or less
dimensions are invariant under a  discrete duality group $G({\bf
Z})$ then the generic functions of the scalars that arise in
    the action transform as   automorphic forms. We then give a
construction of such automorphic forms
    and find that they involve the weights of $G$. As a result,  the
occurrence of weights in the dimensional reduction of M theory can
be thought of as a consequence of the presence of an underlying
    discrete duality group $G({\bf Z})$ of the string theory in lower
dimensions and so interpreted as evidence for such a symmetry.

The  systematic method of constructing  automorphic forms that we
present relies on the ability to construct a  non-linear
realisation,  $\varphi$ from  linear representation $\psi$ of the
continuous group $G$. This construction involves the coset
representatives $g(\xi)$   of $  G/H$, where the $\xi$ labels the
coset.  In the conventions of [39], these are  parameterized by
$$
g (\xi)=e^{\sum_{\vec \alpha
 >0}E_{\vec\alpha} \chi_{\vec\alpha} }e^{-{{1}\over {\sqrt{2}}}\vec
\phi\cdot \vec H} \eqno(1.1)
$$
where $\vec H$ comprises the Cartan subalgebra, $E_{\vec\alpha}$ are
the generators associated to the positive roots.    The automorphic
forms, which are generally  non-holomorphic, are  essentially
functions of $\varphi (\xi)$  summed over the representation $\psi$
from  which they are constructed. One finds that the automorphic
forms contain  $g (\xi)$ acting on the representation $\psi$  and so
the  weights of $G$ corresponding to $\psi$ automatically
appear.

The detailed contents of this paper are as follows. In section two
we extend the calculation of the reference [39] to the dimensional
reduction of the perturbative contribution to higher derivative
terms of the IIB string theory effective action.  We will again find
weights of $E_8$. In section three we will examine the consequences
of demanding that the higher derivative corrections of string theory
in any dimension be invariant under $G({\bf Z})$. Such terms contain
functions of the coset fields  $\xi$ that parameterize $G/H$ times
Riemann tensors, field strengths and Cartan forms.  We calculate how
these functions transform under  $G({\bf Z})$ and show that, under
the natural action of the group on the coset variables $\xi$, they
are ``rotated" by  matrices which belong to a representation of $H$.
In section four we begin by showing that these transformations are
precisely those of non-holomorphic automorphic forms of $G({\bf Z})$
which depend on $\xi$. We then give a method of constructing
automorphic forms once we choose a group $G$ together with a
subgroup $H$ and a linear representation $\psi$ of $G$. In
particular, the automorphic form is constructed from the non-linear
representation of $G$ with local subgroup $H$ formed from the linear
representation $\psi$ of $G$. As explained above the group element
of equation (1.1) enters in this process and in this way the
automorphic form will depend on the coset of $G/H$. As a result of
this construction we show that these automorphic form contain the
weights of $G$ associated with the representation $\psi$ and in
particular the dominant term in the limit of small couplings is of
the form $Z_s\sim e^{\sqrt{2}s{\vec w}\cdot {\vec \phi}}$ where
$\vec w$ is a weight of $G$. In section five we provide some
concluding remarks. Appendices A,B and C give some details and
conventions on non-linear and induced representations, group
representations and examples of $SL(n)$ automorphic forms,
respectively.

\bigskip {\large {2. Type IIB Higher Derivative Corrections and Their
Reduction}} \bigskip

In this section we will evaluate the dimensional reduction to three
dimensions of  the  higher derivative terms that appear in type IIB
string theory. Some of these higher derivative terms in ten
dimensions  involving $D^{2k}R^4$ have been discussed in  detail in
[27-33].  In particular we will determine vectors $\vec w$ that
appear in the dimensional reduction as coefficients of the scalar
fields $\vec\phi$  through the factors $e^{\sqrt{2}\vec w\cdot
\vec\phi}$. This is an extension of the calculation that we
performed in [39] for M-theory and more details may be found there,
although here we will use a slightly more efficient method that we
will explain. The higher derivative corrections  in the ten
dimensional IIB theory already include automorphic forms of
$SL(2,{\bf R})$ however  we will only include in our calculations
the perturbative contribution to the automorphic form.  We will find
that the general higher derivative correction leads to vectors $\vec
w$ that  are weights of $E_8$ (more precisely, in the conventions of
[39], these are half-weights of $E_8$).

Since we are going to use a slightly more streamlined method
compared to that used in reference [39] it will be useful to first
consider the dimensional reduction of a generic theory  possessing
two or more spacetime derivatives involving gravity, gauge fields
and scalars on a $n$-torus. Our compactification ansatz  is given by
$$
d\hat s^2 = e^{2\alpha \rho}ds^2 + e^{2\beta\rho}G_{ij} (dx^i +
A^i_\mu dx^\mu) (dx^j + A^j_\mu dx^\mu) \eqno(2.1)
$$
where
$$
\alpha  = \sqrt{{n}\over 2(n+1)}\ ,\qquad \beta = -{\alpha\over n}
\eqno(2.2)
$$
which ensures that we remain in Einstein frame in three-dimensions.
Here $G_{ij} = e_i^{\ \overline k}e_j^{\ \overline l}\delta_{kl}$
and  $e_i^{\ \overline k}$ is a vielbein with $\det e =1$. We adopt
the convention that $i,j,k,\ldots $ world indices and  $\overline
i,\overline  j,\overline  k,\ldots $ are tangent indices. We note
that this ansatz treats all the directions of the torus on the same
footing and as discussed in reference [39], we will be able to carry
out the dimensional reduction so that the $SL(n,{\bf R})$ invariance
is manifest. In particular,   the degrees of freedom of gravity
associated with the torus, apart from any graviphotons enter the
lower dimensional theory through a non-linear realization of
$SL(n,{\bf R})$ with local subgroup $SO(n)$, {\it i.e.} via the
group element
$$
e(\xi) = e^{\sum_{\underline \alpha
 >0}E_{\underline\alpha} \chi_{\underline\alpha} }e^{-{{1}\over {\sqrt
{2}}}\underline\phi\cdot \underline H} \eqno(2.3)
$$
where $\underline H$ forms the Cartan subalgebra, $E_{\underline
\alpha}$ are positive root generators (when $\underline\alpha>0$) of
$SL(n, {\bf R})$ respectively and $\xi$ collectively denotes the
fields $ \chi_{\underline\alpha} $ and $\underline\phi $.  In fact
the terms which contain $e(\xi)$ alone are  built out of the Cartan
forms $e^{-1}\partial_\mu e =S_\mu+Q_\mu$, where $S_\mu$ and $Q_\mu$
are symmetric and anti-symmetric in $\overline i$ and $\overline j$
respectively. As this belongs to the Lie algebra of $SL(n,{\bf R})$
it does not matter which representation for the generators one takes
to evaluate it.

However, the explicit components  of the vielbein, $e_i^{\ \overline
k}$, associated with the torus reduction are given by taking the
generators to be in the fundamental representation, with highest
weight $\underline \lambda^{n-1}$, where $\underline\lambda^i$,
$i=1,...,n-1$, are the fundamental weights of $SL(n,{\bf R})$. We
now explain why this is the case. Given a linear realization of
$SL(n,{\bf R})$ on a vector space whose vectors have the  components
$\psi_a$  we can construct a non-linear realization with components
$\varphi_a$ by\footnote{*}{\small For further discussion of this
construction we refer the reader to appendix A.}
$$
\varphi_a (\xi)= D(e(\xi)^{-1})_a{}^b \psi_b\ , \ \ {\rm or \
equivalently }\ \  |\varphi (\xi) >=U(e(\xi))|\psi> \eqno(2.4)
$$
where $U(e(\xi))$ indicates  the action of the generators on the
vector space to which $|\psi> = \psi_a |e^a>$ belongs.  From
equation (2.4) we see that $\varphi_a(\xi)$ transforms under
$SL(n,{\bf R})$ by transforming the parameters of the coset $\xi$ in
the usual way and by an $SO(n,{\bf R})$ rotation that acts on the
index $a$. In particular if we take $|\psi> = \psi_i|i,\underline
\lambda^{n-1}>$ to be the representation of $SL(n,{\bf R})$, whose
highest weight is $\underline\lambda^{n-1}$, then
$\varphi_{\overline i}(\xi)$ will transform as a vector with respect
to this $SO(n,{\bf R})$ rotation. However the inverse vielbein
$(e^{-1})_{\overline i}{}^j$ converts world indices to tangent
indices and hence converts quantities that transform under
$SL(n,{\bf R})$ into those that transform under $SO(n,{\bf R})$. As
such we may identify
$$
(e^{-1})_{\overline i}{}^j= D(e^{-1}(\xi))_{i}^{\ j}\ . \eqno(2.5)
$$
Acting on a state $|\psi>=\psi_i |i,\underline \lambda^{n-1}>$ with
$U(e(\xi))$ we find that $e_{\overline i}{}^j$ factors of
$e^{-{{1}\over {\sqrt{2}}}\underline\phi\cdot [\underline
\lambda^{n-1}]}$ where $ [\underline\lambda^{n-1}]$  denotes one of
the weights in the $ \underline \lambda^{n-1}$ representation. The
lowest weight in the $ \underline \lambda^{n-1}$ is just the weight
$-\underline \lambda^{1}$ and so we may rewrite this factor as
$e^{{{1}\over {\sqrt{2}}}\underline\phi\cdot [\underline
\lambda^1]}$. Thus we find that $e_{ i}{}^{\overline j}$ contains
factors of $e^{-{{1}\over {\sqrt{2}}}\underline\phi\cdot [\underline
\lambda_1]}$.

The dimensionally reduced theory will involve corrections that
contain field strengths of the form  $F_{\mu_1\ldots \mu_p i_1\ldots
i_k}$, where $i_1, \ldots $ are  worldvolume indices of the torus.
The field strength may also carry other internal indices that we
neglect for the moment, but we will discuss them below. We can
always use the inverse vielbein $e_{\overline i}^{\ j}$ to convert
all worldvolume indices to tangent space indices. Following the same
argument we used to the vielbein given above, this can be viewed as
the conversion of the linear rank $k$ antisymmetric representation
of $SL(n,{\bf R})$ into a non-linear representation whose indices
rotate under $SO(n)$. Consequently,
    $F_{\mu_1\ldots \mu_p \overline i_1\ldots\overline i_k}$ has  a
dependence on the metric of the torus that is equivalent to  acting
with $U(e(\xi)^{-1})$, on the states $|[{\underline \lambda^{n-k}}]>$
where $[{\underline \lambda^{n-k}}]$ are weights in the representation
with highest weight $ \underline\lambda^{n-k}$. Therefore one finds that
the fields $\underline \phi$ associated with the Cartan subalgebra
of $SL(n,{\bf R})$ occur in   $F_{\mu_1\ldots \mu_p \overline
i_1\ldots\overline i_k}$  through the factor $e^{{1\over
\sqrt{2}}\underline\phi\cdot [\underline \lambda^{k}]}$. We recall
here that the weights $|[{\underline \lambda^{n-k}}]>$ include the
highest weight $\underline \lambda^{n-k}$, but also the lowest weight
which is  $- \underline \lambda^{k}$.

Thus the  action after the dimensional reduction contains terms
which involve $e(\xi)$  alone and are constructed from
$e(\xi)^{-1}\partial_\mu e(\xi)$ (and hence is independent of the
representation used) and field strengths, including those generated
from the Riemann tensor, which are taken to have tangent space
indices. In this way the three-dimensional effective action can be
constructed from various building blocks where each one has indices
that transform under $SO(n)$. Invariants are constructed using the
invariant tensor $\delta_{\overline i\overline j}$. Consequently, to
compute the dependence of the final action on $\underline \phi $ one
just has to add up the contributions from each building block.

One also finds factors of $e^{\sqrt 2 \rho}$ which are readily
computed explicitly from the occurrence of the vielbeins using  the
metric ansatz of equation (2.1) as was done in reference [39].

It is also possible to treat any coset symmetries of the original
theory in a similar way to  the $SL(n,{\bf R})$ associated with the
torus. We illustrate this for  the case of the  $SL(2,{\bf R})$
symmetry of the IIB theory [4], as this is the case of most interest
to us here, but the technique is quite general. Type IIB theory
possess two scalars $\chi$ and $\phi$ which belong to the coset
space $SL(2,{\bf R})/SO (2,{\bf R})$. We may choose our coset
representatives of $SL(2,{\bf R})/SO(2)$ as
$$
g(\tau) = e^{E \chi }e^{-{1\over \sqrt{2}}\phi  H}  \eqno(2.6)
$$
where $E$ and $H$ are the  positive root and Cartan subalgebra
generators of $SL(2,{\bf R})$ respectively. It will be useful to
define $\tau = \chi+ie^{-\phi}$;  as $\tau$  undergoes fractional
linear transformations under the action of $SL(2,{\bf R})$ on this
coset. It also contains two three-form field   strengths
$F_{\mu_1\mu_2\mu_3}^a=3\partial _{[\mu_1}A_{\mu_2\mu_3]}$, $a=1,2$.
The gauge fields must transorm as a linear representation of
$SL(2,{\bf R})$, otherwise, if the gauge fields  transformed as a
non-linear representation of $SL(2,{\bf R})$, the composite nature
of the $SO(2)$ matrix would not preserve the form of the  field
strength and this in turn would not maintain gauge invariance.
Therefore the two three-form field strengths $F_{\mu_1\mu_2\mu_3}^a$
must transform in the doublet representation of $SL(2,{\bf R})$.
However, given the field strength $F_{\mu_1\mu_2\mu_3}^a$ we can
convert it into a three-form $G_{\mu_1\mu_2\mu_3}^a$ that transforms
as a non-linear realization of  $SL(2,{\bf R})$ using equation (A.9)
and the action of $U(g(\tau) ^{-1})$. In particular for the doublet
representation the group element of equation (2.6) can be written as
$$
U(g(\tau)) ={1\over\sqrt{{\rm Im}\tau}}\left(\matrix{{{\rm
Im}\tau}&{\rm Re}\tau\cr 0&1\cr}\right) \eqno(2.7)
$$
so that
$$
\eqalign{ G^1_{\mu_1\mu_2\mu_3} &={1\over\sqrt{{\rm Im}\tau}}(
F^1_{\mu_1\mu_2\mu_3} - {\rm Re}\tau F^2_{\mu_1\mu_2\mu_3}) \cr
G^2_{\mu_1\mu_2\mu_3} &=\sqrt{{\rm Im}\tau}F^2_{\mu_1\mu_2\mu_3}\cr
}\eqno(2.8)
$$
and hence we can form the complex combination [4]
$$\eqalign{
G_{\mu_1\mu_2\mu_3} &= G^1_{\mu_1\mu_2\mu_3}
-iG_{\mu_1\mu_2\mu_3}^2\cr &={1\over\sqrt{{\rm
Im}\tau}}(F^1_{\mu_1\mu_2\mu_3} - \tau F^2_{\mu_1\mu_2\mu_3})
}\eqno(2.9) .
$$  The advantage of working with
$G^a_{\mu_1\mu_2\mu_3}$ rather than $F^a_{\mu_1\mu_2\mu_3}$ is that
it is simpler to form invariants since they rotate on their $a$
indices as a vector of $SO(2)$. As a result for every factor of $
G_{\mu_1\mu_2\mu_3}^a$  that occurs one finds a corresponding factor
of $e^{{{1}\over {\sqrt{2}}}\phi [\mu]}$,  where
$$
[\mu]
=\{{1\over \sqrt{2}},-{1\over\sqrt{2}}\} \eqno(2.10)
$$
are the weights that appear in the fundamental representation of
$SL(2,{\bf R})$.

The above technique also applies to  fields that arise from
dualization. The computation of the $\rho$ dependence is
straightforward and is as explained in [39]. The dualization process
changes the position of indices, such as   world volume indices,
from being upper indices to lower indices and visa-versa.  However,
one can apply the above procedure  to the  field after dualization
and read off the resulting dependence on $\underline \phi$. For
example when reducing the Riemann tensor one finds graviphoton field
strengths which carry a single upper $i$ index. After dualisation
these become a scalar fields with a single lower $i$ index and
therefore one finds factors of the form $e^{{1\over
\sqrt{2}}\underline\phi\cdot[\underline \lambda_1]}$.

Let us now apply this method to the IIB theory dimensionally reduced
on a seven torus to three dimensions. We start by giving the form of
the ten-dimensional type IIB effective action which has a manifest
$SL(2,{\bf R})$ symmetry. We will not need to be concerned with
Fermions or exact coefficients. In Einstein frame we have
$$
S= \int d^{10}x\sqrt{- \hat g} \left(\hat R - (\partial\phi)^2 -
e^{2\phi}(\partial \chi)^2 -G^{a\mu_1\mu_2\mu_3}
G_{\mu_1\mu_2\mu_3}^a -  G_{\mu_1\ldots \mu_5} G^{\mu_1\ldots
\mu_5}\right) \eqno(2.11)
$$
The curvature $ R$ and five-form field strength are singlets of
$SL(2,{\bf R})$. As the five-form field strength is self-dual, this
condition must be imposed by hand and so the above action only has a
limited validity but it is sufficient for our current purposes. The
hat on $\hat R$ indicates that it is the Riemann tensor of the full
higher dimensional metric $\hat g$.

We are interested in the dependence on the scalars  $ \phi,\rho$ and
$\underline \phi$ which we assemble into the 8-vector
$$
\vec \phi = (\phi,\rho,\underline\phi) \eqno(2.12)
$$
In three dimensions, after the appropriate dualizations, we only
have scalars. In addition to $\vec \phi $ there are scalars which
arise as gauge fields with all internal indices or through dualizing
one-form gauge fields in three dimensions. We denote all these
additional scalars by $\chi_{\vec \alpha}$. The action will contain
various terms involving derivatives of these scalars along with a
coefficient of the form $e^{\sqrt{2}\vec w\cdot \vec \phi}$ for some
8-vector $\vec w$:
$$
\vec w = (w,\kappa,\underline w) \eqno(2.13)
$$
The first entry $w$ arises from the behaviour of the fields under
the $SL(2,{\bf R})$. The second entry simply records the power of
$e^{\sqrt{2}\rho}$ that accompanies a field after dimensional
reduction. The third component $\underline w$ corresponds to the
$SL(7,{\bf R})$ representation  of the fields.

It will be instructive to first derive the $E_8$ symmetry that
arises when  IIB supergravity is dimensionally reduced to three
dimensions, that is the  reduction of the action of equation (2.11).
The reduction of the Einstein-Hilbert  term $ R$ gives vectors of
the form (see [39])
$$
{\vec w} =(0,0,[\underline\theta])\qquad {\vec w}
=(0,\sqrt{2}{{8}\over {7}}\alpha,[\underline\lambda^1]) \eqno(2.14)
$$
where $\underline\theta = \underline\lambda^1+\underline\lambda^6$
is the highest weight of the adjoint representation of $SL(7,{\bf
R})$ and $[\underline\theta]$ denotes any element in the set of
weights that appear in the adjoint representation, {\it i.e.} the
roots of $SL(7,{\bf R})$. Similarly $[\underline\lambda^1]$ are the
set weights that appear in the fundamental representation of
$SL(7,{\bf R})$, {\it i.e.} $[\underline\lambda^1]=
\{\underline\lambda^1,...,-\underline\lambda^6\}$. This last set of
vector arises from the graviphotons that have been dualized and  the
steps leading to the  $[\underline\lambda^1]$ part of the vector
were outlined explained above.

Next we can consider the dimensional reduction of
$$
\sqrt{- \hat g} G_{\mu \overline i_1\overline i_2}^a  G^{\mu b}_{\
\overline j_1\overline j_2}\delta^{\overline i_1\overline j_1}
\delta^{\overline i_2\overline j_2}\delta_{ab} \eqno(2.15)
$$
The vectors $\underline \phi$ that this term contributes are readily
found using the discussion above.  One finds that as $G_{\mu
\overline i_1\overline i_2}^a$ has two $SL(7,{\bf R})$ indices
associated with its $SL(7,{\bf R})$ transformation and hence one
finds the contribution $[\underline\lambda^2]$ to the vector. Since
it only has one index associated with its $SL(2,{\bf R})$
transformation this leads to a contribution $[\mu]$ to he part of
the vector corresponding to $\phi$. Thus one finds that this term
gives rise to the series of vectors
$$
\vec w=([\mu],{{2\sqrt{2}}\over {7}}\alpha,[\underline\lambda^2])
\eqno(2.16)
$$
One such vector is
$$
\vec\alpha_7 =(-{{1}\over {\sqrt{2}}},{{2\sqrt{2}}\over
{7}}\alpha,-\underline\lambda^5) \eqno(2.17)
$$
Lastly we reduce the ten-dimensional axion term $e^{2\phi}(\partial
a)^2$ which leads directly to
$$
\vec\alpha_8 =(\sqrt{2},0,0) \eqno(2.18)
$$

One can readily verify that $\vec \alpha_i
=(0,0,\underline\alpha_i)$ with $i=1,...,6$, $\vec \alpha_7$ and
$\vec\alpha_8$  are the simple roots of $E_8$ with the corresponding
Dynkin diagram
$$
\matrix{ & & & & & & & &\bullet&\vec\alpha_8 &\cr & & & & & & & & |&
&\cr & & & & & & & &\bullet&\vec\alpha_7&\cr & & & & & & & &|& &\cr
\bullet&-&\bullet&-&\bullet&-&\bullet&-&\bullet&-&\bullet&\cr
\vec\alpha_1& &\vec\alpha_2& &\vec\alpha_3& &\vec\alpha_4& &
\vec\alpha_5& &\vec\alpha_6\cr}
$$
The bottom line contains the $SL(7,{\bf R})$ subalgebra associated
to diffeomorphisms of the torus ({\it i.e.} the gravity line). The
reduction also leads to terms in three dimensions with other vectors
$\vec w$, in particular one must reduce the five-form field
strength. However the remaining vectors one finds turn out to be
non-simple roots of $E_8$. This appearance of this Dynkin diagram
for type IIB string theory has an elegant origin in terms of
$E_{11}$ [62]. This viewpoint   allows one to understand in an
immediate way how the $E_8$ algebra arises in the dimensional
reduction from the fields of the IIB theory.

Let us now consider the reduction of the possible  higher derivative
terms that can arise in the IIB string theory. We first consider the
terms that involve the polynomials in the Riemann tensor multiplied
by a functions of the scalar fields $\tau,\bar\tau$ which have the
generic form (in Einstein frame)
$$
S_E = \int d^{10} x \sqrt{- \hat g} (\hat R)^{{{l\over 2}}}
Z_x(\tau,\bar\tau) \eqno(2.19)
$$
We will take  $Z_x$  to behave as a sum of terms of the form $
e^{-x\phi}$. In fact   $Z_x$ is a non-holomorphic automorphic form
and only its leading order terms, corresponding to string
perturbation theory, behave in this manner as $\phi \to -\infty$. We
will not consider the non-perturbative contributions in the
calculation in this section. The vectors $\vec w$ that arise from
this term are (see [39])
$$
\vec w = \left(-{{x}\over {\sqrt{2}}},\sqrt{2}\left(1-{l\over
2}+{8\over 7}t\right)\alpha ,s[\underline \theta]+t[\underline
\lambda^1]\right) \eqno(2.20)
$$
where $s,t$ are positive integers with $s+t\le {l\over 2}$.  In
particular $2s$ and  $2t$ count the number of $S_\mu$ and
graviphoton field strengths that are contained in the dimensionally
reduced term respectively. To evaluate whether or not these vectors
are weights of $E_8$ we must show that $\vec\alpha_i \cdot \vec w$
is an integer for all $i=1,...,8$. Calculating away gives
$$
\eqalign{ \vec\alpha_i\cdot \vec w&= s[\underline\theta]\cdot
\underline\alpha_i+t[\underline \lambda^1]\cdot
\underline\alpha_i\cr &=m \cr\vec\alpha_7\cdot \vec w&={x\over
2}+{4\over 7} \left(1-{l\over 2}+{8\over 7}t\right)\alpha^2 -
s[\underline\theta]\cdot \lambda^5 -t[\underline\lambda^1]\cdot
\lambda^5\cr &={x\over 2}+{1\over 4} \left(1-{l\over 2}+{8\over
7}t\right)-t{2\over 7} +n \cr&={x\over 2} + {1\over 4}- {l\over
8}+n\cr \vec\alpha_8\cdot \vec w&=-x\cr} \eqno(2.21)
$$ where
$n,m\in{\bf Z}$. The first expression is automatically an integer
because $[\underline \theta]$ is a root and $[\underline \lambda^1]$
a weight of $SL(7,{\bf R})$. In the second expression we have used
the facts that $[\underline\lambda^1] = \underline\lambda^1 -
\underline\alpha $ where $\underline\alpha$ is a positive root of
$SL(7,{\bf R})$ and $\underline\lambda^i\cdot \underline\lambda^j =
{i(7-j)\over {7}}$ for $i<j$.

It is instructive to  transform this term to string frame by
rescaling $g_{\mu\nu} \to e^{-{1\over 2}\phi} g_{\mu\nu}$. This
results in the term
$$
\eqalign{ S_S &= \int d^{10} x \sqrt{-\hat g} e^{({{l}\over
{4}}-{{5}\over {2}})\phi}(\hat R)^{{{l}\over {2}}} Z(\tau,\bar\tau)
\cr &\sim \int d^{10} x \sqrt{-\hat g} e^{({{l}\over {4}}-{{5}\over
{2}}-x)\phi}( \hat R)^{{{l}\over {2}}} \cr} \eqno(2.22)
$$
If  this term is to arise in string perturbation theory then we
require that  ${{l}\over {4}}-{{5}\over {2}}-x = 2g-2$ for some
$g=0,1,2,...$. Thus we find that ${x\over 2} + {{1}\over {4} }-
{{l}\over {8}} = -g$ and hence in this case
$$
\vec\alpha_i\cdot \vec w \in {\bf Z}\ ,\qquad \vec\alpha_7\cdot \vec
w = -g+n \in {\bf Z}\ ,\qquad \vec\alpha_8\cdot \vec w = -x
\eqno(2.23)
$$
Note that there is no condition that $x \in {\bf Z}$. Rather the
condition ${l\over 4}-{5\over 2}-x = -2g-2$ for some $g=0,1,2,...$
only implies that $x$ is a half integer. In reference [27-33],  IIB
higher derivative terms of the  form $D^{2k}\hat R^4$ have been
computed.  For our purposes they  are equivalent to $\hat R^{4+k}$.
In particular for $4+k={l\over 2}=4$ one finds perturbative
corrections at tree level and one loop which have $x=3/2$ while for
$4+k={l\over 2}=6$ one finds perturbative corrections at tree level
and two loops which have $x=5/2$. Thus one indeed finds for these
and the other known cases that $x$ is half integer.

More generally we can consider a term of the form (in Einstein
frame)
$$
S_E = \int d^{10} x \sqrt{-\hat g} (\hat R)^{{l_1\over 2} }(
G^a_{\mu ij})^{l_2}(  F_{\mu ijkl})^{l_3}Z (\tau,\bar\tau)
\eqno(2.24)
$$
where $Z_x (\tau,\bar\tau)$ is a similar function to that used above
and in particular has the same generic $\phi$ dependence, {\it i.e.}
sum of terms of the form $ e^{-x\phi}$ in the perturbative limit.
Using the analysis of [39], or the quicker method explained above,
we can read off the vectors in as
$$
\vec w = \left(-{{x}\over
\sqrt{2}}+[\mu]l_2,\sqrt{2}\left(1-{{l_1}\over {2}}+{8\over
7}t-{5\over 7}{l_2\over 2} -{3\over 7}{l_3\over 2}\right)\alpha
,s[\underline \theta]+t[\underline \lambda^1] +{l_2\over
2}[\underline \lambda_2]+{{l_3}\over {2}}[\underline
\lambda_4]\right) \eqno(2.25)
$$
The only non-trivial tests that this is a weight come from $\vec
\alpha_7\cdot \vec w$ and $\vec \alpha_8\cdot \vec w$. In the later
case we simply have $\vec \alpha_8\cdot \vec w  = -x\pm l_2\in {\bf
Z}$ whereas
$$
\eqalign{ \vec \alpha_7\cdot \vec w &={x\over 2}+{l_2\over 2}
+{1\over 4}\left(1-{l_1\over 2}+{8\over 7}t-{5\over 7}{l_2\over 2}
-{3\over 7}{l_3\over 2}\right)
    -{2\over 7}t - {2\over 7}l_2 - {4\over 7}l_3+n\cr
&={x\over 2}+{l_2\over 2} +{1\over 4}-{l_1\over 8}-{3l_2\over
8}-{5l_3\over 8} +n\cr} \eqno(2.26)
$$
with $n \in {\bf Z}$. Again converting to string frame, where the
dilaton appears through the factor $e^{2(g-1)\phi}$, tells us that
$$
2g-2 = -{x}-{{5\over 2}} -{l_1\over 4}-{3l_2\over 4}-{5l_3\over 4}
\eqno(2.27)
$$
and hence
$$
\vec \alpha_7\cdot \vec w  = -g+n+{l_2\over 2} \in {\bf Z}
\eqno(2.28)
$$
since  $l_2$ must be even. Here we again see that we find weights if
$x \in {\bf Z}$ but generically $x$ is half an integer.

We note that there are more terms that can be considered. For
example, we could include terms involving higher powers of
$\partial\phi$ and $\partial \chi $ however these will behave in a
similar way to $\partial\rho$ which arises from dimensional
reduction of the Riemann tensor. Other terms arise from components
of $G^a_{\mu\nu i}$ and $F_{\mu\nu ijk}$ with two spacetime indices
in three dimensions. These require dualization into scalar fields
but this is complicated by the dilaton (just as was encountered for
the Bosonic string in [39]) however we do not expect that these
terms will alter the conclusion.

In this section we have examined the possible  higher derivative
corrections that  can arise in the IIB string theory. We have
computed the vectors $\vec w$  associated with the scalars $\vec
\phi = (\phi,\rho,\underline\phi)$. For the lowest order terms of
IIB supergravity itself these belong  to the root lattice of $E_8$,
in fact they are positive roots of the  adjoint representation of
$E_8$. The dilaton dependence is constrained by demanding that the
terms arise as a perturbative correction of IIB string theory.
Requiring that this is the case one finds that the vectors $\vec w$
are half-weights of $E_8$, using the conventions of [39]. Although
we note here that the vectors $\vec w$ are weights with respect to
$SL(7,{\bf R})$, they are only half-weights with respect to the
$8$-th node of $E_8$ which is associated with the $SL(2,{\bf R})$
symmetry of IIB supergravity.

\bigskip {{\large 3 Automorphic Forms in Higher Derivative
Corrections}} \bigskip

As mentioned above,  the  IIA and IIB supergravity theories encode
all the low energy effects of IIA and IIB string theories and so
must contain all non-perturbative low energy effects including
phenomena which are not calculable from our known formulations of
string theory. One of the most interesting properties of   the IIB
supergravity theory is that it possesses an $SL(2,{\bf R})$ symmetry
[4]. Furthermore,  if one dimensionally reduced either  the IIA or
the IIB supergravity theories on a $n-1$ torus, or the eleven
dimensional supergravity theory on an $n$ torus,  one finds the same
set of supergravity theories and remarkably these possess an $E_n$
symmetry for $n\le 9$ [8,9,10].

The dimensionally reduced maximal supergravity theories on a torus
are also the low energy effective actions for the the type II string
theories on an $(n-1)$ torus, or the ill understood M-theory on a
$n$ torus. As we already mentioned they are invariant under a
continuous symmetry group which is non-linearly realized with
respect to a local subgroup. It will prove useful to describe the
representations of this symmetry that the fields in these theories
belong to and as this discussion applies to many such theories we
will denote the non-linearly realised group by $G$ and the local
subgroup by $H$. For the IIB theory $G=SL(2,{\bf R})$ and the local
subgroup is $SO(2)$, whereas when dimensionally reduced to four
dimensions one finds $G=E_7$ and $H=SU(8)$  and  $G=E_8$ and
$H=SO(16)$ in three dimensions. It turns out that in all the cases
we will consider the local subgroup $H$ is just the Cartan
involution invariant subgroup. We recall that the Cartan involution
$I$ is an automorphism, {\it i.e.} it obeys $I(g_1g_2)=I(g_1)I(g_2)
\forall g_1,g_2\in G$, such that $I^2=1$ and acts on the Chevalley
generators as $I(H_a)=-H_a$, $I(E_a)=-F_a,\ I(F_a)=-E_a$.

If one dimensionally reduces to three dimensions one finds, using
suitable dualizations, a theory with just scalars which belong to
the coset $G/H$. In this paper we will work with the coset
representatives that we denote by $g(\xi)$. These transform under a
rigid transformation  $g_0\in G$ as $g(\xi)\to g(\xi')$ where
$$
g_0g(\xi)=g(\xi')  h(g_0,\xi) \eqno(3.1)
$$
and  $h(g_0,\xi)\in H$ is the compensating transformation required
to restore the choice of coset representative. This induces a
non-linear realization of $G$ on the parameters $\xi$ which we
denote by $\xi' = g_0\cdot \xi$.

The dynamics of the scalars is constructed from the Cartan form
$g^{-1}\partial_\mu g$ which takes values in the Lie algebra of $G$
and is invariant under the rigid transformations $g(x)\to g_0g(x)$.
The Cartan form can be written as
$$
g^{-1}\partial_\mu g=P_\mu+Q_\mu \eqno(3.2)
$$
where $Q_\mu$ is in the Lie algebra of $H$.  Our choice of local
subgroup $H$ is odd under the Cartan involution $I$ ($I(h)=-h$ for
$h\in H$)  and so $I(Q_\mu)=Q_\mu$ and then
$P_\mu=g^{-1}\partial_\mu g-I(g^{-1}\partial_\mu g)$ and so
satisfies $I(P_\mu)= -P_\mu$. This implies  that the commutators of
generators of the Lie algebra of $H$ with the generators which are
odd under the Cartan involution leads to generators which are also
odd. As such, under a the local transformation $g(x)\to g(x) h(x)$
we find $P_\mu\to h^{-1}P_\mu h$, while $Q_\mu$ transforms as
$Q_\mu\to h^{-1}Q_\mu h+h^{-1}\partial_\mu h$. The invariant low
energy Lagrangian for the scalars is then given by ${\rm Tr} (P_\mu
P^\mu )$.

If one dimensionally reduces on a torus to a dimension above three
then one will find Bosonic fields other than scalars, in particular
in addition to gravity one will find gauge fields. As we discussed
in the last section, any gauge fields must transform linearly under
the rigid transformations $g_0$ of the group $G$ (see (A.8));
$$
U(g_0)\psi_a = D(g_0^{-1})_a{}^b\psi_b \eqno(3.3)
$$
Consequently the field strengths also  transform as in equation
(3.3). However, as explained at the end of appendix A, using the
scalar fields of the theory, we can always convert a field that
transforms under a linear representations of $G$ into field that
transforms under the non-linear representation
$$
U(g_0)\varphi_a(\xi) = D(h^{-1}(g_0,\xi))_a{}^b\varphi_b(\xi)
\eqno(3.4)
$$
by taking $\varphi_a(\xi)=D(g^{-1}(\xi))_a{}^b\psi_b$. To respect
gauge invariance we must perform  this conversion on the field
strength and not on the gauge fields.

The scalars by themselves always occur with their derivatives as in equation
(3.2). However the quantity $Q_\mu$ only occurs in the dynamics as a
connection for spacetime derivatives acting on fields, such as field
strengths, leaving the  scalars to appear through $P_\mu$. The
Fermions also transform as a non-linear realization. Therefore, all
the fields that appear in the dynamics of IIB supergravity theory
and IIB supergravity  dimensional reduction on a $(n-1)$ torus (or
equivalently the IIA supergravity theory on a $n-1$ torus or
M-theory on an $n$-torus) can be taken to transform as a non-linear
representation of $G$ with local subgroup $H$, {\it i.e.} as in
equation (3.4) for some representation $D$ of $H$.

As mentioned above the continuous groups $SL(2,{\bf R})$ and  $E_n$
are symmetries of the IIB supergravity theory and this theory
dimensionally reduced on an $(n-1)$ torus respectively. Although
these theories are the low energy effective actions for the type IIA
and IIB string theories on an $n-1$ torus, these continuous
symmetries are not symmetries of the underlying  string theories or
M-theory. The supergravity theories possess solitonic solutions
corresponding to strings and branes and the symmetries rotate the
field strengths and charges associated with these solitons. However,
the latter are subject to quantization conditions [21] and has been
conjectured in string theory that the symmetries survive if these
groups were restricted to a discrete subgroup which preserves the
lattice of charges [22,23,24].  The precise form of this group is
clear for the IIB theory; it is just the one generated by two by two
matrices with integer entries and determinant one. This is the so
called U-duality conjecture  which can be thought of as a
combination of the T-duality, which is known to be a valid symmetry
of string theory, combined with the $SL(2,{\bf Z})$ symmetry of the
IIB theory.

In this section we will consider the higher derivative corrections
that can occur in string theories where some of the dimensions are
tori. We will assume that they are invariant under  the discrete
group $G({\bf Z})$  mentioned above and our aim is to discover what
are the consequences of demanding such symmetries on the general
form of such corrections. We will also assume that the fields in
effective actions of such theories transform in the same way that
they did in the low energy effective action. In effect this assumes
that there is a choice of field variables such that the
transformation rules are unaffected by higher derivative terms. That
is the fields occur in expressions which involve their spacetime
derivatives and transform as  in equation (3.3), except that now the
rigid $g_0$ transformations will belong to $G({\bf Z})$ rather than
the continuous group $G$.  When expressed in Einstein frame the
higher derivative terms  are of the generic form
$$
\int d^d x\sqrt{-g} Z(\xi) X \eqno(3.5)
$$
where $X$ is a polynomial in the Riemann curvature,  the modified
field strengths and the covariant derivatives of scalar fields. All
these quantities   will transform as in equation (3.3). An important
exception to the above statement is the appearance of the function
$Z(\xi)$ of the scalar fields $\xi$ which belong to the coset space
$G/H$. Such a function does not contain spacetime derivatives and
their appearance signals  the fact that we no longer have invariance
under the continuous $G$ symmetry, but only under its discrete
subgroup $G ({\bf Z})$.

Since the objects that make up $X$ transform as in equation (3.4),
it follows that $X$ itself, will transform as
$$
U(g_0)X=D(h^{-1}(g_0,\xi))X \eqno(3.6)
$$ where  $g_0$ is a
transformation of  $G({\bf Z})$ and $h(g_0,\xi)$ is the compensating
$H$ transformation required  in equation (3.1), that is $g(\xi)\to
g_0g(\xi)=g(\xi')h(g_0,\xi)$ and for suitable representation $D$.
Demanding that the higher derivative term be invariant under $G({\bf
Z})$ we find that
$$
Z(g_0\cdot\xi)=D(h(g_0,\xi)) Z(\xi) \eqno(3.7)
$$
When carrying out the variation it is important to note that
$Z(\xi)$ is an explicit function of $\xi$ and so its variation just
changes the value of $\xi$ under the action of $g_0\in G({\bf Z})$.
As we will explain in the next section, this last equation is just
the transformation property of an automorphic form of $G({\bf Z})$.
We note that these automorphic forms are not in general holomorphic
and indeed are in most cases non-holomorphic. In section three we
will also discuss the additional constraints, such as differential
equations as well as growth conditions, that non-holmorphic
automorphic forms are expected to also obey.

The simplest case is when $X$ is invariant  under the
transformations of $G({\bf Z})$, {\it i.e.} $D=1$. It is then
obvious that $Z(\xi)$ is inert and so $Z(g_0\cdot\xi)=Z(\xi)$. Such
is the case if $X$ is a polynomial in the Riemannn tensor,  or when
spacetime derivatives act on a polynomial of Riemann tensors. Such
examples have been studied in detail for the IIB theory in reference
[27-36].

Thus we conclude that if we assume that the higher derivative
corrections are invariant under a $G({\bf Z})$ duality symmetry then
every possible term will generically contain functions of the scalar
fields, which belong to the coset space $G/H$, that transform as
automorphic forms of the group $G({\bf Z})$.

We now  illustrate the above discussion in the familiar context of
the IIB string theory as this will allow us to make contact with the
work of references [27-38]. In the previous section we discussed the
$SL(2,{\bf R})$ formulation of the IIB supergravity theory where the
local subgroup is $SO(2)$. As explicitly derived in appendix $C$,
under an element
$$
g_0 = \left(\matrix{a&b\cr c&d\cr} \right)\in SL(2,{\bf Z})
\eqno(3.8)
$$
the compensating $SO(2)$ transformation is given by
$$
h= \left(\matrix{\cos\theta_c &  -\sin\theta_c\cr \sin\theta_c&
\cos\theta_c\cr}\right) ,\qquad e^{2i\theta_c}= {c\tau+d\over c
\bar\tau+d} \eqno(3.9)
$$
All of the type IIB fields  transforms  as equation (3.4) with $D$
which is given by
$$
D(h^{-1})=e^{-i{q}\theta_c} \eqno(3.10)
$$
for some $q$. In particular $q=0$ for the metric and five-form
whereas $q=1,-1$ for the three-form $G_{\mu_1\mu_2\mu_3}$ and its
complex conjugate respectively. The two scalars  belong to the coset
$ SL(2,{\bf R})/SO(2) $. The Cartan involution odd part of the
Cartan forms, $P_\mu$ transform under $ SL(2,{\bf R})$ by a matrix
which is in the doublet representation of $SO(2)$ which is
reducible. Writing this representation as $P_\mu^a $, the two
irreducible representations are given by
    ${\cal P}_\mu= P_\mu^1+iP_\mu^2$  and its complex conjugate $\bar
{\cal P}_\mu= P_\mu^1-iP_\mu^2$ with $q=2,-2$ respectively.

The above  discussion on higher derivative terms is easy to apply to
the IIB supergravity theory.  The object $X$ of equation (3.4) will
have a total charge $q_T$ which is just the sum of the charges of
its factors.  The corresponding automorphic form $Z(\xi)$ that
multiplies $X$ must transform as
$$
Z_{X}(g_0\cdot\tau) =e^{iq_T\theta_c}Z_X(\tau) \eqno(3.11)
$$
This agrees with the discussions of a number of terms given in
reference [27-38].

We can also apply it to higher derivative terms of the IIB theory
involving derivatives of scalars. Such a term, which involves only
the scalars,   will be of the form
$$
\int d^{10} x\sqrt{-g} Z_{r,s}(\tau,\bar \tau) {\cal P}_\mu^r \bar
{\cal P}^{\mu s} \eqno(3.12)
$$
As ${\cal P}^r_\mu \bar{\cal P}^{\mu s}$ has a total $U(1)$ weight
$2(r-s)$ we find that $Z$ transforms as
$$
Z_{r,s}(g_0\cdot\tau) =\left({c\tau+d\over c
\bar\tau+d}\right)^{r-s}Z_{r,s}(\tau) \eqno(3.13)
$$
It is obvious how to generalize this discussion to terms that
involve derivatives of the scalars as well as other objects.

As another example, let us consider the higher derivative terms of
superstring theory on a seven torus or M-theory on an eight torus.
The only dynamical Bosonic fields of the low energy theory are
scalars and they possess an $E_8$ symmetry with local subgroup
$SO(16)$. As explained above, the scalars arise in the dynamics of
the low energy effective action through the Cartan forms of equation
(3.2) which belong to the Lie algebra of $E_8$ and so are in the
$248$ dimensional adjoint representation. The $Q_\mu$ which occurs
in this equation belongs to the Lie algebra of $SO(16)$ which is the
$120$ dimensional adjoint representation. Therefore the $P_\mu$
belongs to a $128$ dimensional representation of $SO(16)$ and must
be a Majorana-Weyl spinor $P_{\mu\alpha}$ in sixteen dimensions. The
kinetic term for the scalars arises in the low energy effective
action as $\bar P_\mu P^\mu$, the bar now being the Majorana
conjugate and we have  suppressed the spinor index. The higher
derivative corrections are of the form of equation (3.5) where $X$
is a polynomial of $P_{\mu \alpha}$ which transforms as in equation
(3.4) where the specific representation matrix $D$ of $SO(16)$
depends on how the polynomial is constructed. The automorphic form
$Z(\xi)$ of the $128$ scalars $\xi$ will therefore transform as in
equation (3.7) with the same $D$. To be concrete  consider  the
higher order term with $2r$ spacetime derivatives that contains a
term of the form
$$
\int d^3 x\sqrt{-g} (\bar P_{\mu_1} \gamma^{a_1}P^{\mu_1})\ldots
(\bar P_{\mu_r} \gamma^{a_r}P^{\mu_r}) Z_{a_1\ldots a_r}(\xi)
\eqno(3.14)
$$
It follows that $Z_{a_1\ldots a_r}(\xi)$ is an automorphic form of
$E_8$ that transforms with a matrix $D$  that is in the rank $r$
symmetric tensor representation of  $SO(16)$ and whose argument that
is the $SO(16)$ compensating transformation.

\bigskip
{\large 4 Automorphic Forms and Induced Representations}
\bigskip

In this section we will show that automorphic forms arise naturally
from the theory of induced representations. As a consquence of
adopting this view point we will find that  they have precisely the
same transformations  as do  the functions of the scalar fields
$Z(\xi)$ that occur in the higher derivative corrections in equation
(3.6). In this way will be able to identify the  $Z(\xi)$ factors as
having the transformation properties of automorphic forms. We will
also give a procedure for constructing automorphic forms for a
general group $G$ with local subgroup $H$. Much of the mathematics
literature on automorphic forms is restricted to the particular case
of $SL(2,{\bf R})$ with  local subgroup $SO(2)$. In this section we
will give a limited account of automorphic forms which we expect
will cover all the possibilities that occur in the higher derivative
corrections of string theory and M-theory.  Automorphic forms for
higher derivative corrections were also discussed in [37], include
their relation to string theory. In particular explicit examples
were given for the cases of $SL(n,{\bf R})/SO(n,{\bf R})$,
$SO(d,d)/SO(d)\times SO(d)$ and $E_d/H$ and second order
differential equations which these automorphic forms satisfy were
given. The examples given in [37] can be constructed by the method
that we give below. Another discussion of automorphic forms intended
for physicists is given in [63].

For a group $G$ with local subgroup $H$ we consider the coset space
$G/H$ whose coset representative are denoted by $g(\xi)$. The group
$G$ has natural action on the coset and therefore also  on the coset
representatives which transform under transformations $g_0\in G$
through equation (3.1).  This coset space  will be of dimension
dim$G$-dim$H$ and in general this will not be an even number,  as
is, for example, the case for $G=SL(n)$ and $H=SO(n)$ if $n\equiv
0,3\ {\rm mod} 4$. Therefore the coset space does not in general
have a complex structure and even when it does we will consider
non-holomorphic automorphic forms. For the application we have in
mind in this paper the coset labels $\xi$ are scalar fields and will
depend on spacetime. However this will play no role in the
considerations in this section, indeed the dependence of $\xi$ on
spacetime is always the same in all equations.

We consider an induced representation of a group $G$ with local
subgroup $H$ which consists of map $\Phi$ from  the coset  $G/H$ to
a vector space $V$ that has the transformation rule ({\it c.f.}
equation (A.4))
$$
U(g_0)\Phi_a(\xi)= D(h^{-1}(g_0,\xi))_a{}^b\Phi_b(g_0\cdot\xi)
\eqno(4.1)
$$ where
$D$ is a linear realization of $H$,  and $h(g_0,\xi)$ is the
compensation of equation (3.1).

Rather than considering  the continuous group $G$ to act on $G/H$ we
now replace this action by that of   a discrete  $G({\bf Z})$. For
example, instead of $SL(n,{\bf R})$ we can consider the discrete
group defined from its fundamental representation with integer
entries, that is we consider the group of $n \times n$ matrices with
integer entries with determinant one. We then consider functions
$\Phi$ which transform as in equation (4.1), but  now with $g_0 \in
G({\bf Z})$. We note that although $\Phi$ transforms under the
discrete group $G({\bf Z})$ it depends on the coset $G/H$ associated
with the continuous group.

Automorphic forms of $G({\bf Z})$ arise from induced representations
if we demand that $\Phi$ is invariant under the action of $G({\bf
Z})$ ;
$$
U(g_0)\Phi_a(\xi)=\Phi_a(\xi) \eqno(4.2)
$$
It then follows that
$$
\Phi_a(g_0\cdot\xi )= D(h(g_0,\xi))_a{}^b\Phi_b(\xi) \eqno(4.3)
$$
The simplest case is when $D(h(g_0,\xi))$ is the identity matrix, in
which case  the index $a$ takes only one value and  the automorphic
form is simply invariant. Although this may not be familiar in this
form, it is the transformation of a automorphic form. In appendix C
we will show that it does indeed agree  with the familiar results
for the much studied case of $SL(2,{\bf Z})$. Imposing equation
(4.3) for the continuous group would of course mean that $\Phi_a$ is
a constant as any two points on the coset are related by a group
element of $G$. However, this is not the case for the discrete group
whose fundamental domain is the coset $G({\bf Z})\setminus G/H$.

The transformation of equation (4.3) is the same as the
transformation  of the coefficients $Z(\xi)$  which  appear in the
higher derivative terms of string theory  discussed in  section
three. This  followed by demanding that these higher derivative
terms be $G({\bf Z})$ invariant.  Therefore we can identify the
coefficients $Z$ as automorphic forms.  However, as we are dealing
with non-holomorphic modular forms they should also satisfy some
additional   conditions, such as differential equations, which we
will discuss later  in this section.

So far we have defined an automorphic form on the coset $G/H$,
however, one can also define them on the group by taking functions
$\Phi_L$ from the group to the vector space $V$ which are induced
representations in the sense of equation (A.4) under the discrete
group $G({\bf Z})$, but also satisfy $U(g_0)\Phi_L(g)=\Phi_L(g)$.

To continue it is useful to compare our treatment of automorphic
forms with that which is usually encoutered in the mathematics
literature for the case of $SL(2,{\bf Z})/SO(2)$. The transformation
of the automorphic form $\Phi$ is often written  as
$$
\Phi(g_0\cdot\xi )= J(g_0,\xi)\Phi(\xi) \eqno(4.4)
$$
where $J(g_0,\xi)$ is called the automorphy factor. This latter
factor is usually just a function, but more generally it is a matrix
acting on $\Phi$ with elements that depend on $g_0$ and $\xi$.
Evaluating $\Phi(g_0'\cdot (g_0\cdot\xi) )=\Phi((g_0'g_0)\cdot\xi)$
we conclude that
$$
J(g_0'g_0,\xi)=J(g_0',g_0\cdot\xi)J(g_0,\xi) \eqno(4.5)
$$
which is consistent with identifying the factors $D(h(g_0,\xi))$ as
automorphy factors as a consequence of equation (A.3).

We now construct some automorphic forms from a linear irreducible
representation $R$, with  components $\psi_a$, of the group $G$.
Given  any such representation $R$ we can form a non-linear
representation with components $\varphi_a (\xi)$ which depends on
the coset $G/H$ by taking ({\it c.f.} equation (A.9))
$$
\varphi_a (\xi)= D(g^{-1}(\xi))_a{}^b\psi_b \eqno(4.6)
$$
We are interested in the restriction of this representation to the
subgroup $G({\bf Z})$ under which the components $\varphi_a (\xi)$
transforms under  $g_0\in G({\bf Z})$ as  (see  equation (A.10))
$$\eqalign{
U(g_0)\varphi_a(\xi) &=
D(g^{-1}(\xi))_a{}^bD(g_0^{-1})_b{}^c\psi_c\cr &= D((g_0
g)^{-1}(\xi))_a{}^b\psi_b \cr &= D(h^{-1}(g_0,\xi))_a{}^b\varphi_b
(\xi') }\eqno(4.7)
$$
Although we started with an irreducible representation of $G$ it
will not be an irreducible representation of $G({\bf Z})$. To obtain
an irreducible representation we restrict our states to a discrete
lattice $\Lambda_R\subset V$. To construct $\Lambda_R$ one can take
a fixed basis of $V$ and then act on it with $G({\bf Z})$.

The automorphic forms are essentially functions of the non-linear
representation $\varphi_a(\xi)$ averaged over the representation
$\psi_a$ from which it is constructed, that is functions of the
generic form
$$
\Phi(\xi)=\sum_{\Lambda_R}f(\varphi_a(\xi))=
\sum_{\Lambda_R}f(D(g^{-1}(\xi))_a{}^b\psi_b) \eqno(4.8)
$$
where $f:V \rightarrow V'$ is a function into some vector space $V'$
and we have suppressed any indices on $\Phi(\xi) $ and $f$.  The sum
is over the lattice $\Lambda_R$ which are the states in the discrete
representation $R$.

Let us first  construct automorphic forms that are invariant under
$G({\bf Z})$ and so consider taking  $f$ of the form
$$
f(\varphi_a(\xi))= K(u(\xi))\equiv f(\xi), \eqno(4.9)
$$
for some function $K:{\bf C} \rightarrow {\bf C}$. Here  $ u(\xi)$
is constructed from the dual and Cartan involution twisted
representations introduced in equations (B.9) and (B.11). In
particular, we   take $ u(\xi)$ to be given by
$$
u(\xi)\equiv  \varphi_{I D }(\xi)^a \varphi_a(\xi) =\psi_{I D}^a
D({\cal M}^{-1}(\xi))_a{}^b\psi_b \eqno(4.10)
$$
where ${\cal M}(\xi)=g(\xi)g^\#(\xi)$. The automorphic form of
equation (4.8) is given by
$$
\Phi(\xi)=\sum_{\Lambda_R}  K(u(\xi)) \eqno(4.11)
$$

Using equation (4.8) and equations (B.14) and (B.16) we find that
under the transformation  $g_0\in G({\bf Z})$ that $\varphi_{I D}^a
(\xi)$ transforms as
$$
U(g_0)\varphi_{I D}^a (\xi)=\varphi_{ I D}^b (\xi')
D(h(g_0,\xi))_b{}^a \eqno(4.12)
$$
It is clear  from equations (4.7) and (4.11) that
$$
U(g_0) K(u(\xi))=K(u(\xi'))\ \ {\rm and\ \  so }\ \ U(g_0)
\Phi(\xi)=\Phi(\xi') \eqno(4.13)
$$
We note that $\varphi_{I D}^a (\xi)$ and $\varphi_{ D}^a (\xi) $
transform in the same way as we assumed that the subgroup $H$ is
invariant under the Cartan involution $I$, i.e. $I(h)=h, \forall h\in
H$.
However, had we taken the latter instead of the former then $u(\xi)$
would be independent of $\xi$ and so $K$ would be uninteresting.

Lastly will show that $\Phi(\xi)$ is invariant. We note that
$\Phi(\xi)$ is constructed from $\varphi_a(\xi)$ which is in turn
given by equation (4.6) in terms of $\psi_b$. Examining  the  action
of $U(g_0)$ on $\varphi_a(\xi)$ given in equation (4.7) we see that
its effect can also be viewed as replacing $\psi_b$ by
$D(g_0^{-1})_b{}^c\psi_c$. However, this just rearranges the states
in the lattice $\Lambda_R$ and as we are summing over all states we
conclude that the total is invariant and hence $U(g_0)\Phi(\xi)=
\Phi(\xi)$. Together with equation (4.13) implies that
$$
\Phi(\xi)=  \Phi(\xi') \eqno(4.14)
$$
in other words it transforms as an invariant automorphic form.

A natural   choice of $K(u(\xi))$ is to take
$$
K(u(\xi))= {1\over (u(\xi))^{s}}. \eqno(4.15)
$$
and in appendix C we will show that this choice along with taking
$\psi_a$ to be the vector representation of $SL(2,{\bf R})$  leads
to the invariant non-holomorphic Eisenstein series of $SL(2,{\bf
Z})$.

We now construct automorphic forms that transform in a non-trivial
way under the action of $G({\bf Z})$. let us take
$$
f_a(\xi)= \varphi_a(\xi) K(u(\xi)), \ \ {\rm or \ equivalently } \ \
\Phi_a(\xi)=\sum_{\Lambda_R}\varphi_a(\xi) K(u(\xi)) \eqno(4.16)
$$
We note that $\Phi_a(\xi)$ is a map from $G/H$ to the vector space
$V$ which carries the representation $R$.

Using equations (4.7) and (4.17) we find that $f_a(\xi)$ transforms
under $g_0\in G({\bf Z})$ as
$$
U(g_0) f_a(\xi)= D( h^{-1}(g_0,\xi))_a{}^b f_b(\xi') \eqno(4.17)
$$
Since the matrix factor $D( h^{-1}(g_0,\xi))_a{}^b $ is independent
of what is being summed over it follows  that
$$
U(g_0) \Phi_a(\xi)= D( h^{-1}(g_0,\xi))_a{}^b \Phi_b(\xi')
\eqno(4.18)
$$
Following the same argument as above which interprets this
transformation as a change in the sum over the representation, we
conclude that
$$
\Phi_a(\xi')= D(h(g_0,\xi))_a{}^b  \Phi_b(\xi) \eqno(4.19)
$$
in other words it transforms as  an automorphic form.

The above construction can be generalized in several ways that may
be important for  the automorphic forms that occur in the higher
derivative corrections of string theory. Firstly, one can give a
more general construction of $u(\xi)$. An invariant under the
transformations of $G({\bf Z})$,   apart from the usual
transformation of the coset variables, can be found by taking  any
function of $\varphi_a$ which is invariant under $\varphi_a\to
D(h)_a{}^b\varphi_b$ for all $h\in H$. Although the latter is not a
transformation  of  $G({\bf Z})$, the  invariance of $u(\xi)$ under
it then ensures that $u(\xi)$   is invariant under $G ({\bf Z})$ up
to the usual  transformation of $\xi$.  This is a consequence of the
fact that the composite matrices $D(h^{-1}(g_0,\xi))_b{}^a $ that
arises in the $U(g_0)$ transformation of $u(\xi)$ will cancel out.
As noted elsewhere, for  our special choice of subgroup $H$, there
is a choice of coset representative such that the $U(h_0),\ h_0\in
H$ transformation of $\varphi_b$ will be by a matrix which just
$D(h_0^{-1})_a{}^b$  and $\xi$ will be a linear representation of
$H$.

We may also generalize the construction by considering automorphic
forms which are a lattice sums over $\varphi(\xi)_a \varphi(\xi)_b
K(u(\xi))$, or more general polynomials. The automorphic forms will
then transform by composite matrices belonging to symmetric tensor
products of the $H$-representation that occurs for $\varphi_a$. In
fact we will use this possibility to construct automorphic forms for
$SL(n,{\bf Z})$ in appendix C. One could also use a non-linear
realization that is constructed from a different  linear
representation to $\psi_a$ for the factors that are outside
$K(u(\xi))$.

We note that  the automorphic form is constructed from
$\varphi_a(\xi)$, which, as shown in equation (4.7), has  the usual
transformation of $\xi$ under the action of the group $G({\bf Z})$
as well as a rotation   by a matrix which depends on an,  albeit
composite, element of $H$. As such, the most general construction is
essentially determined by finding  invariants, or other tensors, of
the $H$-representation of $\varphi_a(\xi)$, even though the symmetry
group is  $G({\bf Z})$. The situation has some similarities to the
case of the  construction of non-linear realizations of the
continuous group $G$. These  can be constructed from
$g^{-1}\partial_\mu g$, or more precisely for the case of scalars
alone from $P_\mu= g^{-1}\partial_\mu g-I((g^{-1}\partial_\mu g)$.
This transforms under $G $ as $P_\mu(\xi)\to
h^{-1}(g_0,\xi)P_\mu(\xi')  h(g_0,\xi)$. As a result, $P_\mu(\xi) $
is   just a particular instance of a non-linear representation $
\varphi_a (\xi)$. In general what higher order invariants one can
construct depends on the invariants that exist in the tensor
products of the $H$-representations that occur in $P_\mu(\xi) $.

There is an essential difference between the construction of
non-linear realization and the construction of automorphic forms
which is crucial for this paper. For the continuous groups the
effective action  for the scalars alone is constructed from
$g^{-1}\partial_\mu g$ and this involves the roots of the Lie
algebra. However,  for the discrete group $G({\bf Z})$ we find that
automorphic form depends on the coset fields $\xi$ that are
contained in $g(\xi)$ and which can be chosen to be of the form
$$
g(\xi)= e^{\sum_{\vec \alpha
 >0}E_{\vec\alpha} \chi_{\vec\alpha} }
e^{-{{1}\over {\sqrt{2}}}\vec\phi\cdot \vec H} \eqno(4.20)
$$ where
$\vec H$ are the Cartan sub-algebra generators and $E_{\vec\alpha}$
are the positive root generators of $G$. In fact, the explicit
construction given above actually involves $g(\xi)$ only through
$$
{\cal M}^{-1}=e^{-\sum_{\vec \alpha
 >0}E_{\vec\alpha} \chi_{\vec\alpha} }
e^{{{\sqrt{2}}}\vec\phi\cdot \vec H}e^{\sum_{\vec \alpha
 >0}E_{\vec\alpha} \chi_{\vec\alpha} } \eqno(4.21)
$$
although as discussed more general possibilities may occur. The
fields $\xi$ that parameterize the coset are made up of the fields
associated with the above generators; we will refer to them as the
Cartan sub-algebra fields $\vec\phi$ and the``axions"
$\chi_{\vec\alpha}$ respectively. We will now show that in the
automorphic forms discussed above one finds that weights of $G$,
rather than the roots, appear as the coefficients of the Cartan
subalgebra fields $\vec \phi$.

It is particulary instructive to study the perturbative contribution
to the automorphic form. In addition to the Cartan subalgebra fields
$\vec\phi$ the automorphic form depends on the ``axion" fields
$\chi_{\vec \alpha}$. Within the context of string theory these
modes arise from components of gauge fields (or in type IIB string
theory as Ramond-Ramond $0$-form). As such there is a perturbative
shift symmetry $\chi_{\vec \alpha}\to \chi_{\vec \alpha} +
\epsilon_{\vec\alpha}$ for an arbitrary $\epsilon_{\vec\alpha}$.
These symmetries typically arise from $U(1)$ gauge transformations
that are not single valued on the torus. In the full quantum theory
the holonomy of a $U(1)$ gauge field around a circle is required to
vanish so that the wavefunction is single valued. The allowed gauge
transformations are therefore restricted and one finds that the
continuous shift symmetry is broken to a discrete one. This implies
that the corresponding scalar field is periodic. However this
discreteness cannot be seen in a perturbative calculation where the
gauge fields are taken to be small fluctuations about the trivial
configuration. Thus the ``axions" only occur in the non-perturbative
contributions to the automorphic form. In fact, the automorphic
forms have a sort of periodicity under integer shift in
$\chi_\alpha$ and so possess a Fourier expansion in $\chi_\alpha$.

Since,  the perturbative contribution  is independent of
$\chi_\alpha $, we  can find this contribution by first  setting
$\chi_\alpha=0$ and then taking the  perturbative limit. In other
words, the perturbative part of the automorphic form can be
calculated by first restricting  $g(\xi)$ to its Cartan subalgebra
and then taking the perturbative limit. Thus we make the replacement
$$
g(\xi)\to h(\vec \phi)= e^{-{{1}\over {\sqrt{2}}}\vec\phi\cdot \vec
H} \eqno(4.22)
$$
We note that in this case ${\cal M}\to e^{ {\sqrt{2}}\vec\phi\cdot
\vec H}$ and as a result, we find that
$$
u(\xi)\to  <\psi_{I D}|U({\cal M}^{-1}) |\psi> =
e^{\sqrt{2}\vec\phi\cdot [\vec \Lambda]} <\psi_{I
D}|\psi>\eqno(4.23)
$$
where $\underline\Lambda$ is the highest weight of the
representation $|\psi>$. In order for the lattice sum to converge it
must be that $K(u)\to 0 $ as $u\to \infty$ so let us assume that, at
large $u$, $K=u^{-s}$ with $s>0$. In the perturbative limit the
lattice sum will be dominated by states for which $\vec\phi\cdot
[\vec \Lambda]$ is the most negative\footnote{*}{\small It is
possible that more than one weight will contribute but we will
ignore this issue here.}
$$
\eqalign{ K &\to \sum_{\Lambda_R}{e^{-\sqrt{2}s\vec\phi\cdot [\vec
\Lambda]}\over <\psi_{I D}|\psi>^s} \cr&\sim
e^{-\sqrt{2}s\vec\phi\cdot {\vec w}_\Lambda}
\sum_{\Lambda_R'}{1\over <\psi_{I D}|\psi>^s} \cr &\sim
N_se^{-\sqrt{2}s\vec\phi\cdot{\vec w}_\Lambda}\cr}\eqno(4.24)
$$
where ${\vec w}_\Lambda$ is the weight in  the representation of
$[\vec\Lambda]$ for which $\vec\phi\cdot [\vec \Lambda]\to -\infty$
the most quickly, $\Lambda_R'$ is the set of states in $\Lambda_R$
with this weight and $N_s=\sum_{\Lambda'_R}<\psi_{I D}|\psi>^{-s}$
is a constant.

Lastly we must consider the contribution of $\varphi_a$ in equation
(4.16) for the cases where the automorphic form has a non-trivial
transformation under $H$. In the limit that we can set the
``axions'' to zero we have that
$$\eqalign{
\varphi_a &= D(g^{-1})_a^{\ b}\psi_b \cr &\to D\left(e^{{{1}\over
{\sqrt{2}}}\vec\phi\cdot \vec H}\right)_a^{\ b}\psi_b\cr &=
e^{{{1}\over {\sqrt{2}}}\vec\phi\cdot \vec
w_\lambda}\psi_a}\eqno{(4.25)}
$$
which has the same form as (4.24). Thus we see that, in the
perturbative limit, $\phi_a\sim e^{-\sqrt{2}s'\vec\phi\cdot {\vec
w}_\Lambda}$ and hence we find weights or half-weights if $s' \in
{\bf Z}$ or $s' \in {\bf Z}+{1\over 2}$ respectively.

Even for a given theory there are several ways to take the
perturbative limit, depending on which of the components of
$\vec\phi$ associated with the Cartan subalgebra we choose to take
to $-\infty$. Typically on expects that each component can be
associated to some coupling constant or physical parameter. For
example in section two we saw that the physical radius of the torus
is proportional to $e_i{}^{\overline i}\sim e^{-{\beta}\rho }
e^{-{1\over \sqrt 2}[\underline \lambda_1]\cdot \underline \phi}$
thus there will be various limits corresponding to which radii
become large. Depending on which component of $\vec \phi$ one takes
large one finds that different weights in $\Lambda_R$ lead to the
dominant behaviour in the limit. To give an explicit example, we
consider the type IIB string theory on a seven torus, the
perturbative limit associated with the string coupling in ten
dimensions consists of taking the dilaton $\phi\to -\infty$ large.
This  implies that the volume modulus $\rho$ and the torus `shape'
moduli $\underline \phi$ can be kept finite. The explicit form for
the roots of $E_8$ were given in section 2 we find the fundamental
weights are
$$
\eqalign{ \vec\lambda^i &= (0,2\sqrt{2\over
7}i,\underline\lambda^i)\ \ ,\ i=1,..,5 \cr \vec\lambda^6 &=
(0,5\sqrt{2\over 7} ,\underline\lambda^6)\cr \vec\lambda^7 &=
(0,\sqrt{14} ,\underline 0) \cr \vec\lambda^8 &=
({1\over\sqrt{2}},\sqrt{7\over 2} ,\underline 0) \cr }\eqno(4.26)
$$
The first space in the above vectors correspond to the position of
the dilaton field $\phi$.  We see that in the perturbative limit
only $\vec\lambda^8\cdot \vec\phi\to -\infty$. If we express ${\vec
w}_\Lambda = n_i\vec \lambda^i$ then one sees that that the dominate
term in the expansion of $\phi$ comes from a weight ${\vec
w}_\Lambda$ with the largest non-vanishing value of $n_8 $. In
M-theory the weak coupling limit, in so far as it exists, is where
the curvatures are small. There is no dilaton but instead the volume
modulus must be large, so that $\rho\to -\infty$, with the `shape'
moduli $\underline\phi^i$ fixed. The explicit weights of $E_8$ that
arise from compactification of M-theory (using the anstaz (2.1))
were given in [39] as
$$
\eqalign{ \vec\lambda^i &= ({3\sqrt{2}\over
4}i,\underline\lambda^i)\ \ ,\ i=1,..,4 \cr \vec\lambda^i &=
({5\sqrt{2}\over 4}(8-i) ,\underline\lambda^i)\ \ ,\ i=5,6,7\cr
\vec\lambda^8 &= (2\sqrt{2},\underline 0) \cr }\eqno(4.27)
$$
where $\underline \lambda^i$ are the fundamental weighs of the
$SL(8,{\bf R})$ symmetry associated to the eight torus upon
compactification to three dimensions. In this limit we see that
$\vec \lambda\cdot\vec\phi\to -\infty$ for all the fundamental
weights but does so most quickly for $\vec\lambda^5$.

Let us close this section with some additional comments on
automorphic forms. Unlike holomorphic forms, non-holomorphic forms
are generally specified by more than just their transformation
properties as one cannot use concepts such as analyticity to deduce
the full function from a knowledge of its poles or asymptotic
behaviour. Indeed for the case of $SL(2,{\bf R})$ the
non-holomorphic automorphic forms are usually defined to transform
as in equation (4.3) but also to be an eigenvalue of the $SL(2,{\bf
R})$ invariant Laplacian and  behave as ${\rm Im}\tau\to \infty$
like $\phi(\tau)\sim ({\rm Im}\tau)^{N}$ for some fixed $N$. In
fact, the $SL(2,{\bf R})$ invariant Laplacian is just the Casimir of
$SL (2,{\bf R})$ when the generators  correspond to their natural
action on the coset $SL(2,{\bf R}) /SO(2,{\bf R})$. A similar
picture is true for the case of $SL(3,{\bf R})$ but now the
automorphic forms obey two differential equations; indeed they are
required to be eigenvalues of the two Casimirs of $SL(3,{\bf R})$
[64].

It is natural to consider non-holomorphic automorphic forms of  $G$
to satisfy $r$ differential equations where $r$ is the rank of $G$.
In particular one might  demand that they be eigenvaulues of the $r$
Casimirs of $G$ whose generators are   realized by  their natural
action on the coset   $G/H$. We note  that the perturbative
contribution of the automorphic forms constructed above depend on
$r$ scalar fields associated with the Cartan subalgebra of $G$ and
the values of the $r$ Casimirs will be  given in terms of the
highest weight of the representation used to construct the
automorphic form.  Thus it would seem likely that there is an
alternative way to characterize these automorphic forms by
specifying  their transformation rule, as in equation (4.3) and a
particular  highest weight of the representation.

We note that the situation for the automorphic forms that occur in
the higher derivative corrections is likely to be more complicated.
In particular the invariant automorphic form that occurs for the
$D^6 R^4$ term in the IIB theory [27-33] is not an eigenvalue of the
$SL(2,{\bf R})$ invariant Laplacian, but rather solves the
eigenvalue problem in the presence of sources obtained from other
automorphic forms that appear  at lower order in the effective
action. It would be good to understand these differential equations
more generally, as was done in [29] for type IIB string theory where
they arise as a consequence of the higher order corrections to
supersymmetry and also to understand how such differential equations
might arise naturally from the mathematical viewpoint.

\bigskip {\large {5 Conclusion}}
\bigskip

In this paper we have given a systematic method of constructing
automorphic forms once one specifies a  group $G$ and subgroup $H$,
which we  took  to be the Cartan involution invariant subgroup, as
well as a linear  representation $\psi$ of $G({\bf Z})$. The
automorphic form is built from the non-linear representation
$\varphi$ constructed from $\psi$ which involves the coset
representatives $g (\xi)$ of $G/H$ acting on $\psi$. In this way the
dependence of the automorphic form on the coset $G/H$ appears and it
follows from the construction that  the automorphic forms involves
the weights of $G$ corresponding to the representation $\psi$.

We also showed that if the higher derivative corrections to the type
II strings in any dimension were invariant under a duality group
$G({\bf Z})$ then the  functions of the scalars that occur could, by
considering   their transformation properties,  be identified with
automorphic forms.

Lastly we found that the dimensional reduction of the higher
derivative corrections of the IIB theory to three dimensions on a
torus lead to weights of $E_8$, generalizing the similar result of
[39] for M theory. Since, as we just explained above, the type II
effective actions must involve automorphic forms and so weights if
they are invariant under a $G({\bf Z})$ duality group, we can
interpret the appearance of weights upon dimensional reduction as
evidence for such an underlying duality symmetry of M theory

In closing we note that there is an important difference between
dimensional reduction and compactification. The former discards all
the Kaluza-Klein and wrapped brane modes while the latter keeps
them. In general the dimensional reduction of a higher derivative
term only leads to a part of the corresponding term in the  lower
dimension. In particular it will not lead to an automorphic form of
the full lower dimensional duality group. Rather one can only expect
to find the part of the automorphic form that survives the limit
where the compact directions are taken to infinite radius. On the
other hand one would expect that, given the full higher derivative
term calculated in the compactified theory one can obtain the
correct higher derivative term in the uncompactified theory by
taking the radii to infinity. However  compactification of loop
amplitudes  has been found [27-36] to lead to the full automorphic
forms themselves, at least from eleven to nine dimensions.

It has been observed [65] that since $E_{11}$ involves the
$SL(2,{\bf R})$ symmetry of the IIB theory and this later symmetry
is broken to $SL(2,{\bf Z})$ then $E_{11}$  itself must be broken to
a discrete symmetry. This means, for example,   that even Lorentz
transformations contained in the  $E_{11}$ symmetry are discrete.
This paper presents a first step in how one might implement a
discrete $E_{11}$ symmetry in M theory and indeed what this could
be. One might like to study automorphic forms based on $E_{11}$ and
hope that this would encode all, or a large parts, of the effective
action.

\bigskip
{\large Acknowledgements}
\bigskip

We would like to thank A. Pressley for discussions.  This work was
supported in part  by the PPARC grant PPA/G/O/2000/00451 and PCW is
support by a  PPARC senior fellowship PPA/Y/S/2002/001/44.

\bigskip {\large Appendix A: Non-linear and Induced Representations}
\bigskip

In this appendix we summarize some basic facts about non-linear
representations [66] and induced representations  that will be
needed in this paper. A non-linear realization of a group $G$ with
respect to a subgroup $H$ considers group elements $g\in G$, which
depend on spacetime, and are taken to transform as
$$
g(x)\to g_0g(x)\ \ {\rm and } \ \ g(x)\to g(x)h(x) \eqno(A.1)
$$
where $g_0$ is any element of $G$ and is a rigid transformation,
that is independent of spacetime, and $h(x)$ is an element of $H$
which depends on spacetime and so is a local transformation.  Any
theory invariant under the above two transformations can be thought
of as the non-linear realization of $G$ with respect to $H$. In
general the result will not be unique, but if the action has only
two spacetime derivatives then it is constrained up to just a few
constants. Furthermore if the subgroup $H$ is large enough then the
action will indeed be uniquely determined. We note that in this
section the spacetime dependence of $g$ and $h$ just goes along for
the ride and hence we are just describing the  usual transformations
on the coset space $G/H$ induced by the natural action of the group.

Associated with the second transformation of equation (A.1) we  see
that  invariant quantities of the theory will only depend on the
coset space $G/H$. One can use this transformation to fix a set of
coset representatives $g(\xi)$ where $ \xi$ are the parameters that
label the cosets, {\it i.e.} the equivalence classes. Once one makes
this choice the transformation under $g_0$ will in general no longer
preserve the choice of coset representative and one must make a
compensating $H$ transformation
$$
g(\xi)\to g_0g(\xi)=g(g_0\cdot\xi)h(g_0,\xi) \eqno(A.2)
$$
Here $h(g_0, \xi) $ is the required compensating  transformation,
which was denoted by $h^{-1}$ in reference [39]. We will often
denote the action on the coset coordinates by $\xi\to
\xi'=g_0\cdot\xi$. To simplify the notation we have drop the
explicit spacetime dependence of $\xi$, as it is not relevant in
this mathematical account and as the dependence of $\xi$ on
spacetime is not changed by any of the steps in this appendix.
Evaluating $g_0^1g_0^2g(\xi)$ as $(g_0^1g_0^2)g(\xi)$ or
$g_0^1(g_0^2g(\xi))$ and comparing the two we find the consistency
condition
$$
h(g_0^1g_0^2,\xi)=h(g_0^1,\xi^1)h(g_0^2,\xi) \eqno(A.3)
$$
where $g_0^1g(\xi)=g(\xi^1)h(g_0^1,\xi)$.

For the groups $G$ and subgroups $H$ of interest to us, the Lie
algebra of $G$ can be written as the Lie algebra of $H$ plus an $H$
invariant compliment, denoted $H^\perp$.  This means that the
generators of $H^\perp$ possess  commutators with the elements in
the Lie algebra of $H^\perp$ that are again  in $H^\perp$. This is
guaranteed if the algebra $G$ possess an automorphism which squares
to one such that the generators of $H$    and those of $H^\perp$
transform into themselves with a minus  and plus sign respectively.
For the groups we have in mind the subgroups $H$ are by definition
those that are preserved by the Cartan involution $I$ and as a
result the generators of $H$ and $H^\perp$ transform in the required
way, {\it i.e.} $I(H)=H, I(H^\perp)=-H^\perp$.  In this case the
coset representatives can be chosen to be constructed from the
generators of $H^\perp$ and then they obey $h_0g(\xi)=g(\xi')h_0$
for $h_0\in H$. Consequently  $\xi$ transforms linearly under $H$
and $h(h_0,\xi)=h_0$.

An induced representation of a group $G$ with respect to a subgroup
$H$ consists of a set of functions $\varphi$ which map $G$ to some
vector space $V$  which carries a linear representation $D(h)$ of
$H$ where $h\in H$. If $\varphi_a$ are the components of $\varphi$,
they are required to satisfy the condition
$$
\varphi(gh)_a=D(h^{-1})_a{}^b\varphi_b(g) , \forall g\in G,\ h \in H
\eqno(A.4)
$$
The
transformation of the group $G$  is defined by
$$
U(g_0)\varphi(g)=\varphi(g_0g),\ \forall g, g_0\in G, \eqno(A.5)
$$
In fact $\varphi$ does not really depend  on the full group $G$, but
only on the  coset $G/H$ as by  equation (A.5) the value of
$\varphi$ at  two points in the same coset is the same  up to  the
matrix factor  $D(h^{-1})$.  As such, we can define a function on
the coset $G/H$ by
$$
\varphi_a(\xi)=\varphi_a(g(\xi)) \eqno(A.6)
$$
where $g(\xi)$ are the above discussed coset representatives. The
transformation of equation (A.5) then becomes
$$\eqalign{
U(g_0)\varphi_a(\xi)&=\varphi_a(g_0g(\xi))\cr
&=\varphi_a(g(g_0\cdot\xi)h(g_0,\xi))\cr &=
D(h^{-1}(g_0,\xi))_a{}^b\varphi_b(g_0\cdot\xi) \cr} \eqno({A.7})
$$
One can verify that it is indeed a representation using equation
(A.3). As noted above,  for the subgroups $H$ of interest to us one
can make a choice of coset representative such
$D(h(h_0,\xi))=D(h_0)$ if $h_0\in H$ and so for these
transformations $D(h)$ is independent of $\xi$ and is just the usual
representation matrix and the action of $h_0$ on $\xi$ is just a
linear realization. In this sense $\varphi$ just transforms linearly
under the subgroup $H$.

Given any linear representation of $G$ carried by an element $\psi
\in V$
$$
U(g_0) \psi_b=D(g_0^{-1})_a{}^b \psi_b, \eqno(A.8)
$$
we can convert
it into a non-linear representation of the form discussed above. To
do this we define
$$
\varphi_a(\xi)=D((g^{-1}(\xi))_a{}^b\psi _b \eqno(A.9)
$$ whereupon
it transforms as
$$
\eqalign{
U(g_0)\varphi_a(\xi) &=
D(g^{-1}(\xi))_a{}^bD(g_0^{-1})_b{}^c\psi_c\cr
&=D((g_0g(\xi))^{-1})_a{}^b\psi_b \cr
&=D((g(\xi')h(g_0,\xi))^{-1})_a{}^b\psi_b \cr
&=D(h^{-1}(g_0,\xi))_a{}^b\varphi(\xi')_b\cr} \eqno({A.10})
$$
where $h(g_0,\xi)$ are the $H$ group elements of equation (A.2). We
note that  $\varphi$ transforms  under a representation of $G$, but
the matrix $D$ has an argument that only involves the group element
$h(g_0,\xi)$ which  belongs to $H$.  We will  refer to this as a
non-linear representation.

It can  happen that one finds that $\varphi$ transforms under more
that one irreducible representation of $G$ as the matrix $D$ is not
an irreducible representation of $H$. Nevertheless we find that we
can always convert the linear realization of equation (A.8) to the
non-linear realization of $G$ given in equation (A.8).

In the above we have used the passive interpretation of
transformations. For example, for a linear realisation of equation
(A.8) it means that
$$\eqalign{
U(g_0^1)U(g_0^2)\psi_a &=U(g_0^1)D((g_0^2)^{-1})_a{}^b\psi_b\cr
&=D((g_0^2)^{-1})_a{}^bD((g_0^1)^{-1})_b{}^c\psi_c\cr
&=D((g_0^1g_0^2)^{-1})_a{}^b\psi_b\cr
&=U(g_0^1g_0^2)_a{}^b\psi_b\cr} \eqno(A.11)
$$
{\it i.e.} $U(g_0^1)U(g_0^2)=U(g_0^1g_0^2)$ as it should for a
representation.

\bigskip {\large Appendix B}
\bigskip

In this appendix we will give an account of certain aspects  of the
theory of group representations that are required in this paper.
This appendix is similar to that of reference [39], but we will
explicitly  use the passive interpretation of transformations and
give the expressions in terms of components.  We recall that a
linear representation $R$ of a group $G$ consists of a vector  space
$V$ and a set of operators $U(g),\ \forall g\in G$ which act on $V$,
namely $|\psi> \to U(g)|\psi>$ such that $ U(g_1g_2) =U(g_1)U(g_2)$.
If the vector space has a basis $| e^a >$ we can write
$|\psi>=\psi_a| e^a>$ where we use the  repeated index summation
convention.  The action of the group is given by
$$
U(g)|\psi>=  \psi_a(U(g)| e^a >)= (U(g)\psi_a)| e^a >  \eqno(B.1)
$$
where
$$
U(g)\psi_a= D(g^{-1})_a{}^b\psi_b \ \  {\rm and\ so}\ \  U(g)| e^a
>= | e^b >D(g^{-1})_b{}^a \eqno(B.2)$$ We note that while the
components $\psi_a$  transform with argument $g^{-1}$, the vectors
$| e^a >$ transform with $(g^{-1})^T$

In this paper we will take  the algebra $G$ to be  finite
dimensional semi-simple  and simply laced. The states in the
representation can be chosen so as  to be eigenstates of $\vec H$.
The eigenvalues are called weights. It can be shown that the weights
of $G$ belong to  the dual lattice to the lattice of roots, {\it
i.e.} a weight $\vec w$ satisfies
$$
{\vec w}\cdot{\vec \alpha_a}\in {\bf Z} \eqno(B.3)$$ for the simple
roots  $\vec \alpha_a$. The representations of interest to us are
finite dimensional and so must have a highest weight $\vec \lambda$
which is the one such that $\vec \lambda+\vec \alpha_a$ is not a
weight for all simple roots $\vec \alpha_a$. The representations
will  also have a lowest root denoted $\vec \mu$. Of particular
interest are the fundamental representations which are those whose
highest weights $\vec \lambda^a$ obey the relation
$$
\vec\lambda^a\cdot\vec\alpha_b = \delta^a_{b} \eqno(B.4)$$ for all
simple roots  $\vec \alpha_a$. The roots are themselves weights and
these correspond to the adjoint representation,  whose highest
weight we will denote by $\vec \theta$.

For $SL(n)$, {\it i.e.} $A_{n-1}$, the fundamental weights $\vec
\lambda^a$ satisfy
$$
\vec\lambda^a\cdot\vec \lambda^b = a(n-b)/n \eqno(B.5)$$ for $b \ge
a$. The representation with highest weight $\vec \lambda^{n-k}$ is
realized on a tensor with $k$ totally anti-symmetrized superscript
indices, {\it i.e.} $T^{i_1\ldots i_k}=T^{[i_1\ldots i_k]}$. Using
the group invariant epsilon symbol $\epsilon ^{i_1\ldots i_n} $,
this representation is equivalent to taking a tensor with $n-k$
lowered indices.

Given any simple root one may carry out its Weyl reflection on any
weight
$$
S_\alpha(w)=\vec w-(\vec \alpha\cdot \vec w)\vec \alpha \eqno(B.6)$$
The collection of all such reflections is called the Weyl group and
it can be shown that any member of it can be written in terms of a
product of Weyl reflections in the simple roots. Although the
precise decomposition of a given element of the Weyl group is not
unique its  length is defined to be the smallest number of simple
root reflections required. However,   there does exist a unique Weyl
reflection, denoted $W_0$,  that has the longest length. This
element  obeys $W_0^2=1$, takes the positive simple roots to
negative simple roots and its length is the same as the number of
positive roots.  As a result, $-W_0$ exchanges the positive simple
roots with each other and,  as  Weyl transformations preserve the
scalar product, it must also preserve  the Cartan matrix.
Consequently,  it must lead to  an automorphism of the Dynkin
diagram. Given any representation of $G$ the highest and lowest
weights are related by
$$
\vec \mu
=W_0\vec \lambda \eqno(B.7)
$$ Given the definition of the
fundamental weights and carrying out a Weyl transformation $W_0$, we
may conclude that the negative of the highest and lowest weights of
a given fundamental representation are the lowest  and highest
representation  of one of the other fundamental representations. It
is always the case that the two representations have the same
dimension. However it can happen that a fundamental representation
is self-dual.

For $SL(n)$ $W_0=(S_{\vec\alpha_1}\ldots S_{\vec\alpha_
{n-1}})(S_{\vec\alpha_1}\ldots S_{\vec\alpha_{n-2}}) \ldots
(S_{\vec\alpha_1}S_{\vec\alpha_2})S_{\vec \alpha_1}$ and one finds
that, in this case,
$$
W_0 \vec \lambda _{n-k}=  \vec \mu _{n-k}=- \vec \lambda _{k} \iff
W_0 \vec \mu _{n-k}=  \vec \lambda _{n-k}=- \vec \mu _{k} .
\eqno(B.8)$$ This result also follows from the above remarks on
$W_0$ as it must take a fundamental representation to a fundamental
representation and correspond to an automorphism of the Dynkin
diagram which in this case is just takes the nodes $k$ to $n-k$.

Given a linear representation $R$ acting on $|\psi> \in V$ we may
consider the dual representation $R_D$  that is carried by the space
of linear functionals, denoted  $V^*$,  acting on $V$. The group
action is defined by
$$
<\psi_D | \to <U(g) \psi_D|=<\psi_D |U(g^{-1})\ ,\ \forall g \in G,\
<\psi_D |  \in V^* \eqno(B.9)
$$
We note that $<\psi_D |\psi>$ is $G$-invariant. If we introduce a
dual basis $e_a^*$ for $V^*$ such that $e_a^*(e^b)\equiv
<e_a^*|e^b>=\delta_a^b$ we can express $<\psi_D |=e_a^* \psi_D^a$.
 From the invariance of the scalar product and equation (B.2) we find
that the transformation in terms of the components is given by
$$
\psi_D^a\to U(g) \psi_D^b =\psi_D^a D(g)_a{}^b \eqno(B.10)
$$

Since the linear functionals  carry a representation we may also
choose a basis for them that is  labeled by the weights. It is easy
to see that  a linear functional with a weight $\vec w$ only has a
non-zero result on a state with weight $-\vec w$. A little further
thought allows one to conclude that  if the representation $R $ has
highest and lowest weight $\vec \lambda$ and $\vec \mu$ respectively
then the dual representation has a highest weight $-\vec \mu$ and
lowest weight $-\vec \lambda$. Indeed the dual representation has
the same dimension as the original representation. For the case of
$SL(n)$, {\it i.e.} $A_{n-1}$, if the representation $R$ is the
fundamental representation with highest weight $\vec \lambda^k$ then
it follows from equation (B.9) that the dual representation is the
fundamental representation with highest weight $\vec \lambda^{n-k}$.
Thus the representations carried by $T^{i_1\ldots i_{(n-k)}}$ is
dual to the representation carried by $T^{i_1\ldots i_k} $ or
equivalently carried by $T_{i_1\ldots i_{(n-k)}}$ if we lower the
indices with epsilon.

Given a representation $R$ and any automorphism $\tau$ of the group
$\tau$ ({\it i.e.} $\tau(g_1g_2)=\tau (g_1) \tau(g_2)$) we may also
define a twisted representation $R_\tau$ on the same vector space
$V$ as follows. If $|\psi_\tau>$ are the states of the twisted
representation we may write  $|\psi_\tau>= \psi_{\tau a}| e^a >$
then the components transform as
$$
\psi_{\tau a} \to U(g)\psi_{\tau a}=D(\tau(g^{-1}))_a{}^b \psi_
{\tau b} \ \forall g\  \ \in G. \eqno(B.11)
$$

In this paper   we will take the automorphism to be the Cartan
involution  which we also denoted by $I$. It is easy to see that if
the representation $R$  has highest and lowest  weight $\vec
\lambda$ and $\vec \mu$ respectively then the dual representation
has a highest weight $-\vec \mu$ and lowest weight $-\vec \lambda$
and so the Cartan involution twisted representation is isomorphic to  the dual
representation.

In appendix A we showed, using equation (A.9), how we can convert a
linear representation, with components $\psi_a$, into a non-linear
representation with components $\varphi_a(\xi)$ which transform as
in equation (A.10) under the group element $g_0$ as
$$
U(g_0)\varphi_a(\xi) =D(h^{-1}(g_0,\xi))_a{}^b\varphi(\xi')_b
\eqno(B.12)
$$
Given the dual representation we can also construct an
analogous non-linear representation if   we define the component
fields by
$$
\varphi_D^a(\xi)=\psi_D^b D(g(\xi))_b{}^a \eqno(B.13)
$$
One verifies
that it transforms as
$$
U(g_0)\varphi_D^a (\xi)=\varphi_D^b (\xi') D(h(g_0,\xi)_b{}^a
\eqno(B.14)
$$
Taking the automorphism to be the Cartan involution
$I$ we can similarly construct a non-linear representation from the
twisted linear representation of equation (B.11) by taking the
components
$$
\varphi_{I a}= D(g^\# (\xi) )_a{}^b \psi_{I b} \eqno(B.15)
$$
where  $g^{\#}=(I(g))^{-1}$. This representation transforms as
$$
\varphi_{I a}(\xi)\to U(g_0)(\varphi_{I a}(\xi))=
D(h(g_0,\xi)^{-1})_a{}^b\varphi_{I b}(\xi') \eqno(B.16)
$$
We note
that $h^{\#}=h^{-1}$ as by definition $I(h)=h$.

Examining the above transformations we observe that
$$\eqalign{
\varphi_{I D}^a(\xi) \varphi_a&= \psi_{I D}^b  D(I(g(\xi)))_b{}^a
D((g(\xi)^{-1})_a{}^c\psi_c\cr &=\psi_{I D}^b D({\cal
M}(\xi)^{-1})_b{}^c \psi_c \cr} \eqno(B.17)
$$
where ${\cal M}(\xi)=gg^\#$, is invariant under the non-linear
realization of $G$, using equations (B.12) and (B.14). We note that
the twisted  dual representations $\psi_{I D}^a$ and the original
representation $\psi_a$ are isomorphic to each other. In particular,
for $A_n$ if $\psi_a$ is the representation with highest weight
$\lambda_k$ so is $\psi_{I D}^a$. The expression
$\varphi_{D}^a\varphi_{I a}$ is also invariant under non-linear
transformations of $G$, however, if we consider all representations
of $G$ in the expression  of  equation (B.17) we do not gain any new
invariant quantities by considering this latter quantity.

\bigskip
{\large Appendix C Examples of Automorphic forms of $SL(n,{\bf Z})$}
\bigskip

In section four we have given a general procedure for constructing
automorphic forms which may be unfamiliar to the reader. In this
appendix we apply this formalism first to the case of $SL(2,{\bf
Z})$  and recover some of the well known automorphic forms and then
to the case of $SL(n,{\bf Z})$.

\bigskip {{C.1 $SL(2,{\bf Z})$}}
\bigskip

Let us start by recalling the well known properties of the coset
$SL(2,{\bf R})/SO(2,{\bf R})$. The local subgroup is the Cartan
involution invariant subgroup of $SL(2,{\bf R})$ which is just
$SO(2, {\bf R})$. It consists of the matrices
$$
h(g_0,\tau) = \left(\matrix{\cos\theta&-\sin\theta\cr \sin\theta&
\cos\theta\cr}\right) \eqno(C.1.1)$$ Using such a local
transformation in  equation (A.1) we may choose our coset
representatives $g(\xi) \in SL(2,{\bf R})/SO(2,{\bf R})$  to have
the upper triangular form
$$
g(\chi,\rho) = {{1}\over{\sqrt{\rho}}}\left(\matrix{\rho&\chi\cr 0&
1\cr}\right) \eqno(C.1.2)$$ with $\rho >0$. Thus the pair
$\xi=(\chi,\rho)$ parameterize the coset space $G/H$ and it will be
helpful to introduce $\tau = \chi + i\rho$. Under a discrete
$SL(2,{\bf Z})$ transformation of the form
$$
g_0 =\left(\matrix{a&b\cr c&d\cr}\right) \eqno(C.1.3)$$ one finds
that $g_0g$ is no longer of the form of equation (C.1.1) as it does
not preserve the choice of coset representative. However if we also
consider a local compensating $SO(2,{\bf R})$ transformation as in
equation (A.1) we find that
$$
g_0g(\tau) = g(\tau) h(g_0,\tau)=
{{1}\over {\sqrt{\rho'}}}\left(\matrix{\rho'&\chi'\cr
0& 1\cr}\right) h(g_0,\tau) \eqno(C.1.4)
$$
with
$$
e^{2i\theta}= {c\tau+d\over c\bar \tau +d}
%,\ \ {\rho\over\rho'}= {1\over |(c\bar \tau +d)|^2}
\eqno(C.1.5)
$$
and
$$
\tau'={{a\tau+b}\over {c\tau+d}} \eqno(C.1.6)$$ Note that even
though $g_0$ is a discrete transformation we require $h$ to be a
local transformation since it depends on $\tau$ in addition to
$g_0$. This is the well-known action of $SL(2,{\bf Z})$ on the coset
which one can denote by $ \tau'= g_0\cdot \tau$.

We now construct  automorphic forms using the method given in
section four.  We must choose a representation $\psi$  of
$G=SL(2,{\bf R})$ which we take to be the vector representation.
This is just the column vector $|\psi>=\big (\matrix {\psi_1\cr
\psi_2\cr}\big )$. The dual Cartan involution twisted representation
is just the transpose, that is $<\psi_{I D}|=\big (\matrix {\psi_1 ,
\psi_2\cr}\big)$. Next we consider $G({\bf Z})=SL(2,{\bf Z})$ and to
obtain an irreducible representation we restrict the states to the
lattice $\Lambda_R={\bf Z}^2-\{(0,0)\}$ with elements $|\psi>=\big
(\matrix {m\cr -n\cr}\big ), \ m,n\in {\bf Z}$ and similarly for
$<\psi_{ID}|$. For $SL(n,{\bf R})$ $\#$ is just the transpose and
hence, in the vector representation,
$$
D({\cal M}^{-1})_a{}^b= {1\over \rho}\left(\matrix{1&-\chi\cr -\chi&
\rho^2+ \chi^2\cr}\right) \eqno(C.1.7)
$$
and therefore
$$\eqalign{
u(\tau)&= \psi_{I D}^a D(({\cal M}^{-1})_a{}^b\psi_{b}\cr &=
{|m+n\tau |^2\over Im \tau}\cr} \eqno(C.1.8)
$$
An invariant automorphic form is then given by equation (4.11) with
the choice of $K(u)$ of equation (4.15) and it is given by
$$\eqalign{
\phi(\tau) &=   \sum_{(m,n)\in {\bf Z}^2-\{0,0\}} {1\over
u(\tau)^s}\cr &=\sum_{(m,n)\in {\bf Z}^2-\{0,0\}}{{({\rm
Im}\tau)^s}\over {|m+n\tau|^{2s}}}} \eqno(C.1.9)
$$
We recognize
these as the well known invariant non-holomorphic Eisenstein series.

We now  construct the automorphic forms  that transform
non-trivially. From equation (4.7) we see that
$$
\varphi= {1\over \sqrt \rho}\left(\matrix{m+n\chi\cr
-n\rho\cr}\right) \eqno(C.1.10)
$$ The irreducible representations of
$SL(2,{\bf Z})$ are $\varphi_{\pm}=\varphi_1\pm i\varphi_2$ where
$$
\varphi_{+}={(m+n\bar\tau)\over \sqrt {Im\tau}},\ \ {\rm and }\ \
\varphi_{-}={(m+n \tau)\over \sqrt {Im\tau}} \eqno(C.1.11)
$$ An
automorphic form is given in equation (4.18) and taking into account
the  possible modification discussed below equation (4.21) we
consider
$$\eqalign{
\phi_w(\tau)&=\sum_{\Lambda_R} {(\varphi_{-})^w\over (u(\tau))^s}\cr
&=\sum_{(m,n)\in {\bf Z}^2-\{0,0\}}{{({\rm Im}\tau)^s}\over
{|m+n\tau|^{2s}}}{(m+n\tau)^w\over {Im\tau}^{{w\over 2}}}\cr
&=\sum_{(m,n)\in {\bf Z}^2-\{0,0\}}{{({\rm Im}\tau)^{s-w/2}}\over
{|m+n\tau|^{2s-w}}}\left({m+n\tau\over m+n\bar\tau}\right)^{w\over
2}\cr } \eqno(C.1.12)
$$
It follows from its construction that this
automorphic form transforms non-trivially with a $D(h(g_0,\tau))=
e^{iw\theta_c}$.

Let us now consider the pertubative limit as $\rho\to \infty$. One
readily sees from (C.1.12) that the dominant terms from $n=0$. These
are just the states $|\psi>=\big (\matrix {m\cr 0\cr}\big )$ in the
lattice $\Lambda_R$ with weight $w_{\Lambda} = 1/\sqrt{2}$. Thus we
see that
$$\eqalign{
\phi &\to \sum_{m\in{\bf Z}-0}{\rho^{s-w/2}\over |m|^{2s-w}}\cr &=
2\zeta(2s-w)\rho^{s-w/2}\cr } \eqno(C.1.13)
$$
and indeed we see that this term is independent of $\chi$.

\bigskip
{C.1.2 $SL(n,{\bf Z})$}
\bigskip

Let us now consider automorphic forms for $SL(n,{\bf Z})$. Again we
consider the vector representation and we can generalize the
previous discussion by using the local $SO(n,{\bf R)}$ invariance to
write the coset representatives $g\in SL(n,{\bf R})/SO(n,{\bf R})$
as
$$
g (\rho,\chi) = {1\over (\rho_1\ldots\rho_{n-1})^{{1\over n}}}
\left(\matrix{\rho_1&\rho_2\chi_{12}&\rho_3\chi_{13}&\ldots&\chi_{1n}\cr
&
\rho_2&\rho_3\chi_{23}&\ldots&\chi_{2n}\cr&&\rho_3&\ldots&\chi_{3n}\cr
&&&\ddots&\vdots\cr &&&&1\cr}\right) \eqno(C.2.1)
$$
This is just the product of a matrix involving the $\chi$'s
multiplied by the diagonal matrix ${\rm diag} ({\rho_i})$ which is
of the form of the group element of equation (1.1). One also finds
that the inverse element takes the form
$$
g^{-1} (\rho,\chi) = (\rho_1\ldots\rho_{n-1})^{{1\over n}}
\left(\matrix{\rho_1^{-1}&-\rho_1^{-1}\tilde\chi_{12}&-\rho_1^{-1}\tilde\chi_{13}&\ldots&-\rho_1^{-1}\tilde\chi_{1n}\cr
&
\rho_2^{-1}&-\rho_2^{-1}\tilde\chi_{23}&\ldots&-\rho_2^{-1}\chi_{2n}\cr&&\rho^{-1}_3&\ldots&-\rho_3^{-1}\tilde\chi_{3n}\cr
&&&\ddots&\vdots\cr &&&&1\cr}\right) \eqno(C.2.2)
$$
where $\tilde\chi_{ij}= \chi_{ij} +{\cal O}(\chi^2)$ are polynomials
in $\chi_{ij}$.

Acting with a  discrete $g_0\in SL(n,{\bf Z})$ transformation acting
on $g (\rho,\chi)$ will change this form, however it can then be put
back into upper triangular form by a local $h\in SO(n,{\bf R})$
transformation. This will generate a non-linear realization $\xi\to
g_0\cdot\xi$ where now $\xi$ collectively labels the fields $\rho_i$
and $\chi_{ij}$ for $i<j=1,...,n-1$.

To construct automorphic forms we start with the vector
representation of $SL(n,{\bf R})$ where $|\psi> \in {\bf R}^n$. We
then restrict attention to $SL(n,{\bf Z})$ and take the states
$|\psi> \in \Lambda_R={\bf Z}^n-\{0,...,0\}$. Thus if we take
$$
|\psi> = \left(\matrix{m_1\cr m_2\cr\vdots\cr m_n}\right)
\eqno(C.2.3)
$$
we find that
$$
|\varphi> =(\rho_1\ldots\rho_{n-1})^{{1\over
n}}\left(\matrix{m_1\rho_1^{-1}-m_2\rho_1^{-1}\tilde\chi_{12}-\ldots\cr
m_2\rho_2^{-1}-\ldots\cr\vdots\cr m_n}\right) \eqno(C.2.4)
$$
and
$$
\eqalign{ u(\xi)&= \varphi^a \varphi_{a}\cr &=
(\rho_1\ldots\rho_{n-1})^{{2\over
n}}\left(\rho_1^{-2}(m_1-m_2\tilde\chi_{12}-\ldots)^2+\rho_2^{-2}(m_2-\ldots)^2+...+m_n^2\right)
\cr} \eqno(C.2.5)
$$
We can then find automorphic forms by taking
$$
\Phi(\xi) = \sum_{\vec m\in {\bf Z}^n-{\vec 0}}{1\over (u(\xi))^s}
\eqno(C.2.6)
$$
which are invariant under $SL(n,{\bf Z})$, or
$$
\Phi_{a_1...a_r}(\xi) = \sum_{\vec m\in {\bf Z}^n-{\vec
0}}{\varphi_{a_1}(\xi)...\varphi_{a_r}(\xi)\over (u(\xi))^s}
\eqno(C.2.7)
$$
which will transform in the symmetric $r$-tensor
representation of $SO(n)$ under an $SL(n,{\bf Z})$ transformation.

These expressions are clearly somewhat complicated. However we can
consider the limit where $\rho_i\to 0$. In this case  we find the
automorphic forms are dominated by states with $m_1=...=m_{n-1}=0$
and hence we can set $\chi_{ij}=0$. In this limit we find
$$\eqalign{
\Phi(\xi) &\to (\rho_1\ldots\rho_{n-1})^{-{2s\over
n}}\sum_{m_n\in{\bf Z}-0} {1\over|m_n|^{2s}}\cr
&=2\zeta(2s)(\rho_1\ldots\rho_{n-1})^{-{2s\over n}}}\eqno(C.2.8)
$$
and
$$\eqalign{
\Phi_{a_1...a_r}(\xi) &\to(\rho_1\ldots\rho_{n-1})^{-{2s-r\over
n}}\sum_{m_n\in{\bf Z}-0} {1\over|m_n|^{2s-r}}\cr &=2\zeta(2s-r)
(\rho_1\ldots\rho_{n-1})^{-{2s-r\over n}} } \eqno(C.2.9)
$$

Our last step to show that this limit does indeed have the form of
equation (4.24) in terms of a weight of $SL(n,{\bf R})$. To this end
we consider a decomposition of $SL(n,{\bf R})$ in terms of
$SL(n-1,{\bf R})$. In particular we will work with the fundamental
representation where we can choose a Cartan basis such that
$$
H_i = \left(\matrix{h_i&0\cr0&0\cr}\right), i=1,...,n-2,\qquad
H_{n-1} = {1\over
\sqrt{n^2-n}}\left(\matrix{1&0\cr0&-(n-1)\cr}\right) \eqno{(C.2.10)}
$$
where $h_i$ are the Cartan matrices for $SL(n-1,{\bf R})$. The
generators $E_{\vec\alpha}$ for $\vec \alpha >0$ can then be chosen
to have zeros everywhere except for a single entry above the
diagonal that is equal to one. A straightforward calculation show
that the simple roots take the form
$$
\vec\alpha_i=(\underline\alpha_i,0)\ ,i=1,...,n-2\qquad
\vec\alpha_{n-1} =\left(-\underline\lambda^{n-2},\sqrt{n\over
n-1}\right) \eqno(C.2.11)
$$
where $\underline\alpha_i$ and $\underline\lambda^i$, $i=1,..,n-2$
are the simple roots and fundamental weights of $SL(n-1,{\bf R})$.
The states $|\psi>$ that dominated the sum are of the form
$$
|\psi> =\left(\matrix{0\cr 0\cr\vdots\cr m_n}\right) \eqno(C.2.12)
$$
and hence their $\vec H$ eigenvalue is $\vec w_\Lambda=(\underline
0, -\sqrt{{n-1\over n}})=-\vec\lambda^{n-1}$.  Comparing (C.2.1) and
(1.1) we see that
$$\eqalign{
(\rho_1...\rho_{n-1})^{-{1\over n}}&=\left(e^{\sum_{\vec \alpha
 >0}E_{\vec\alpha} \chi_{\vec\alpha} }e^{-{1\over\sqrt{2}}\vec\phi\cdot\vec H} \right)_{nn} \cr
&=e^{{1\over\sqrt{2}}\sqrt{n-1\over n}\phi_n}\cr
&=e^{-{1\over\sqrt{2}}\vec\phi\cdot \vec\lambda^{n-1}}\cr
}\eqno(C.2.13)
$$
where the subscript $nn$ denotes the $nn$th component of the matrix
representative of $g(\xi)$. Hence we see that, in the limit
$\rho_i\to 0$,
$$
\Phi(\xi)\to 2\zeta(2s)e^{-\sqrt{2}s\vec\phi\cdot
\vec\lambda^{n-1}}\eqno{(C.2.14)}
$$
and
$$\Phi_{a_1...a_r}(\xi)\to
2\zeta(2s-r)e^{-{\sqrt{2}}(s-r/2)\vec\phi\cdot \vec\lambda^{n-1}}
\eqno{(C.2.15)}
$$

\vfil \eject

\bigskip {\bf References}
\bigskip

\item{[1]}
    I.~C.~G.~Campbell and P.~C.~West,
    %``N=2 D = 10 Nonchiral Supergravity And Its Spontaneous
Compactification,''
    Nucl.\ Phys.\ B {\bf 243}, 112 (1984).
    %%CITATION = NUPHA,B243,112;%%

\item{[2]}
    F.~Giani and M.~Pernici,
    %``N=2 Supergravity In Ten-Dimensions,''
    Phys.\ Rev.\ D {\bf 30}, 325 (1984).
    %%CITATION = PHRVA,D30,325;%%

\item{[3]}  M.~Huq and M.~A.~Namazie,
    %``Kaluza-Klein Supergravity In Ten-Dimensions,''
    Class.\ Quant.\ Grav.\  {\bf 2}, 293 (1985)
    [Erratum-ibid.\  {\bf 2}, 597 (1985)].
    %%CITATION = CQGRD,2,293;%%

\item{[4]}
J.~H.~Schwarz and P.~C.~West,
    %``Symmetries And Transformations Of Chiral N=2 D = 10
%Supergravity,''
    Phys.\ Lett.\ B {\bf 126}, 301 (1983).
    %%CITATION = PHLTA,B126,301;%%

\item{[5]}
    P.~S.~Howe and P.~C.~West,
    %``The Complete N=2, D = 10 Supergravity,''
    Nucl.\ Phys.\ B {\bf 238}, 181 (1984).
    %%CITATION = NUPHA,B238,181;%%

\item{[6]}
    J.~H.~Schwarz,
    %``Covariant Field Equations Of Chiral N=2 D = 10 Supergravity,''
    Nucl.\ Phys.\ B {\bf 226}, 269 (1983).
    %%CITATION = NUPHA,B226,269;%%

\item{[7]}
    E.~Cremmer, B.~Julia and J.~Scherk,
    %``Supergravity Theory In 11 Dimensions,''
    Phys.\ Lett.\ B {\bf 76}, 409 (1978).
    %%CITATION = PHLTA,B76,409;%%

\item{[8]}
E.~Cremmer and B.~Julia,
%``The N=8 Supergravity Theory. 1. The Lagrangian,''
Phys.\ Lett.\ B {\bf 80}, 48 (1978).
%%CITATION = PHLTA,B80,48;%%

\item{[9]}
N.~Marcus and J.~H.~Schwarz,
    %``Three-Dimensional Supergravity Theories,''
    Nucl.\ Phys.\ B {\bf 228} (1983) 145.
    %%CITATION = NUPHA,B228,145;%%

\item{[10]}
B.~Julia and H.~Nicolai,
    %``Conformal internal symmetry of 2d sigma-models coupled to
%gravity and  a
    %dilaton,''
    Nucl.\ Phys.\ B {\bf 482}, 431 (1996)
    [arXiv:hep-th/9608082].
    %%CITATION = HEP-TH 9608082;%%

\item{[11]}
B.~Julia, in {\it Vertex Operators and Mathematical Physics},
Publications of the Mathematical Sciences Research Institute no3.
Springer Verlag (1984); in {\it Superspace and Supergravity}, ed.
S.~W.~Hawking and M.~ROcek, Cambridge University Press (1981)

\item{[12]}
S.~Mizoguchi,
    %``E(10) symmetry in one-dimensional supergravity,''
    Nucl.\ Phys.\ B {\bf 528}, 238 (1998)
    [arXiv:hep-th/9703160].
    %%CITATION = HEP-TH 9703160;%%

\item{[13]}
J.~Ehlers, Dissertation, Hamburg University, 1957.

\item{[14]}
    R.~Geroch,
    %``A Method For Generating New Solutions Of Einstein's Equation.2,''
    J.\ Math.\ Phys.\  {\bf 13}, 394 (1972).
    %%CITATION = JMAPA,13,394;%%

\item{[15]}
P.~Breitenlohner, D.~Maison and G.~W.~Gibbons,
    %``FOUR-DIMENSIONAL BLACK HOLES FROM KALUZA-KLEIN THEORIES,''
    Commun.\ Math.\ Phys.\  {\bf 120}, 295 (1988):
    %%CITATION = CMPHA,120,295;%%

\item{[16]}
E.~Cremmer, B.~Julia, H.~Lu and C.~N.~Pope,
    %``Higher-dimensional origin of D = 3 coset symmetries,''
    arXiv:hep-th/9909099.
    %%CITATION = HEP-TH 9909099;%%

\item{[17]}
    N.~D.~Lambert and P.~C.~West,
    %``Coset symmetries in dimensionally reduced bosonic string theory,''
    Nucl.\ Phys.\ B {\bf 615} (2001) 117
    [arXiv:hep-th/0107209].
    %%CITATION = HEP-TH 0107209;%%

\item{[18]}
K.~Kikkawa and M.~Yamasaki,
    %``Casimir Effects In Superstring Theories,''
    Phys.\ Lett.\ B {\bf 149}, 357 (1984);
    %%CITATION = PHLTA,B149,357;%%

\item{[19]}
T.~H.~Buscher,
    %``A Symmetry Of The String Background Field Equations,''
    Phys.\ Lett.\ B {\bf 194}, 59 (1987);
    %%CITATION = PHLTA,B194,59;%%
    %``Path Integral Derivation Of Quantum Duality In Nonlinear Sigma Models,''
    Phys.\ Lett.\ B {\bf 201}, 466 (1988).
    %%CITATION = PHLTA,B201,466;%%

\item{[20]}
    M.~Rocek and E.~P.~Verlinde,
    %``Duality, quotients, and currents,''
    Nucl.\ Phys.\ B {\bf 373}, 630 (1992)
    [arXiv:hep-th/9110053].
    %%CITATION = HEP-TH 9110053;%%

\item{[21]}
M.~Henneaux and C.~Teitelboim,
   %``QUANTIZATION OF TOPOLOGICAL MASS IN THE PRESENCE OF A MAGNETIC POLE,''
   Phys.\ Rev.\ Lett.\  {\bf 56} (1986) 689.
   %%CITATION = PRLTA,56,689;%%

\item{[22]}A.~Sen,
    %``Electric magnetic duality in string theory,''
    Nucl.\ Phys.\ B {\bf 404}, 109 (1993)
    [arXiv:hep-th/9207053].
    %%CITATION = HEP-TH 9207053;%%

\item{[23]}
A.~Font, L.~E.~Ibanez, D.~Lust and F.~Quevedo,
    %``Strong - weak coupling duality and nonperturbative effects in
string theory,''
    Phys.\ Lett.\ B {\bf 249}, 35 (1990).
    %%CITATION = PHLTA,B249,35;%%

\item{[24]}
C.~M.~Hull and P.~K.~Townsend,
    %``Unity of superstring dualities,''
    Nucl.\ Phys.\ B {\bf 438}, 109 (1995)
    [arXiv:hep-th/9410167].
    %%CITATION = HEP-TH 9410167;%%

\item{[25]} K.~A.~Meissner,
    `%`Symmetries of higher-order string gravity actions,''
    Phys.\ Lett.\ B {\bf 392}, 298 (1997)
    [arXiv:hep-th/9610131].
    %%CITATION = HEP-TH 9610131;%%

\item{[26]}
N.~Kaloper and K.~A.~Meissner,
    %``Duality beyond the first loop,''
    Phys.\ Rev.\ D {\bf 56}, 7940 (1997)
    [arXiv:hep-th/9705193].
    %%CITATION = HEP-TH 9705193;%%

\item{[27]}
M.~B.~Green and M.~Gutperle,
    %``Effects of D-instantons,''
    Nucl.\ Phys.\ B {\bf 498}, 195 (1997)
    [arXiv:hep-th/9701093].
    %%CITATION = HEP-TH 9701093;%%

\item{[28]}
    M.~B.~Green, M.~Gutperle and P.~Vanhove,
    %{\sl One loop in eleven dimensions},
    Phys.\ Lett.\ B {\bf 409} (1997) 177
    [arXiv:hep-th/9706175].
    %%CITATION = HEP-TH 9706175;%%

\item{[29]}M.~B.~Green and S.~Sethi,
    %``Supersymmetry constraints on type IIB supergravity,''
    Phys.\ Rev.\ D {\bf 59}, 046006 (1999)
    [arXiv:hep-th/9808061].
    %%CITATION = HEP-TH 9808061;%%

\item{[30]} M.~B.~Green, H.~h.~Kwon and P.~Vanhove,
    %``Two loops in eleven dimensions,''
    Phys.\ Rev.\ D {\bf 61}, 104010 (2000)
    [arXiv:hep-th/9910055].
    %%CITATION = HEP-TH 9910055;%%

\item{[31]}M.~B.~Green and P.~Vanhove,
    %``Duality and higher derivative terms in M theory,''
    JHEP {\bf 0601}, 093 (2006)
    [arXiv:hep-th/0510027].
    %%CITATION = HEP-TH 0510027;%%

\item{[32]} M.~B.~Green, J.~G.~Russo and P.~Vanhove,
    %``Non-renormalisation conditions in type II string theory and
maximal supergravity,''
    arXiv:hep-th/0610299.
    %%CITATION = HEP-TH 0610299;%%

\item{[33]}
A.~Basu,
    %``The D**10 R**4 term in type IIB string theory,''
    arXiv:hep-th/0610335.
    %%CITATION = HEP-TH 0610335;%%

\item{[34]}
J.~A.~Harvey and G.~W.~Moore,
    %``Fivebrane instantons and R**2 couplings in N = 4 string theory,''
    Phys.\ Rev.\ D {\bf 57}, 2323 (1998)
    [arXiv:hep-th/9610237].
    %%CITATION = HEP-TH 9610237;%%

\item{[35]}
E.~Kiritsis and B.~Pioline,
    %``On R**4 threshold corrections in type IIB string theory and
(p,q) string instantons,''
    Nucl.\ Phys.\ B {\bf 508}, 509 (1997)
    [arXiv:hep-th/9707018].
    %%CITATION = HEP-TH 9707018;%%

\item{[36]} B.~Pioline, H.~Nicolai, J.~Plefka and A.~Waldron,
    %``R**4 couplings, the fundamental membrane and exceptional theta
correspondences,''
    JHEP {\bf 0103}, 036 (2001)
    [arXiv:hep-th/0102123].
    %%CITATION = HEP-TH 0102123;%%

\item{[37]}
  N.~A.~Obers and B.~Pioline,
  %``Eisenstein series and string thresholds,''
  Commun.\ Math.\ Phys.\  {\bf 209}, 275 (2000)
  [arXiv:hep-th/9903113].
  %%CITATION = CMPHA,209,275;%%

\item{[38]}
N.~Berkovits and C.~Vafa,
    %``Type IIB R**4 H**(4g-4) conjectures,''
    Nucl.\ Phys.\ B {\bf 533}, 181 (1998)
    [arXiv:hep-th/9803145].
    %%CITATION = HEP-TH 9803145;%%

\item{[39]}
N.~Lambert and P.~West,
%``Enhanced coset symmetries and higher derivative corrections,''
Phys.\ Rev.\ D {\bf 74}, 065002 (2006) [arXiv:hep-th/0603255].
%%CITATION = HEP-TH 0603255;%%

\item{[40]}
    M.~B.~Green and J.~H.~Schwarz,
    %{\sl Supersymmetric Dual String Theory (III). Loops and renormalization}
    Nucl.\ Phys.\ B {\bf 198} (1982) 441.
    %%CITATION = NUPHA,B198,441;%%

\item{[41]}
    D.~J.~Gross and E.~Witten,
    %{\sl Superstring Modifications Of Einstein's Equations},
    Nucl.\ Phys.\ B {\bf 277} (1986) 1.
    %%CITATION = NUPHA,B277,1;%%

\item{[42]}
    M.~T.~Grisaru and D.~Zanon,
    %{\sl Sigma Model Superstring Corrections To The Einstein-Hilbert Action},
    Phys.\ Lett.\ B {\bf 177} (1986) 347.
    %%CITATION = PHLTA,B177,347;%%

\item{[43]}
    N.~Sakai and Y.~Tanii,
    %{\sl One Loop Amplitudes And Effective Action In Superstring Theories},
    Nucl.\ Phys.\ B {\bf 287} (1987) 457.
    %%CITATION = NUPHA,B287,457;%%

\item{[44]}
    R.~C.~Myers,
    %{\sl Superstring Gravity And Black Holes},
    Nucl.\ Phys.\ B {\bf 289} (1987) 701.
    %%CITATION = NUPHA,B289,701;%%

\item{[45]}
D.~J.~Gross and J.~H.~Sloan,
  %``THE QUARTIC EFFECTIVE ACTION FOR THE HETEROTIC STRING,''
  Nucl.\ Phys.\ B {\bf 291}, 41 (1987).
  %%CITATION = NUPHA,B291,41;%%

\item{[46]}
    M.~J.~Duff, J.~T.~Liu and R.~Minasian,
% {\sl Eleven-dimensional origin of string / string duality: A one-loop test},
    Nucl.\ Phys.\ B {\bf 452} (1995) 261
    [arXiv:hep-th/9506126].
    %%CITATION = HEP-TH 9506126;%%

\item{[47]}
    J.~G.~Russo and A.~A.~Tseytlin,
    %{\sl One-loop four-graviton amplitude in eleven-dimensional supergravity},
    Nucl.\ Phys.\ B {\bf 508} (1997) 245
    [arXiv:hep-th/9707134].
    %%CITATION = HEP-TH 9707134;%%

\item{[48]}
M.B.~Green, H.~Kwon and M.~Gutperle,
%``Light-cone quantum mechanics of the eleven-dimensional superparticle,''
  JHEP {\bf 9908}, 012 (1999)
  [arXiv:hep-th/9907155].
  %%CITATION = HEP-TH 9907155;%%

\item{[49]}
S.~Deser and D.~Seminara,
  %``Counterterms/M-theory corrections to D = 11 supergravity,''
  Phys.\ Rev.\ Lett.\  {\bf 82}, 2435 (1999)
  [arXiv:hep-th/9812136];
  %%CITATION = HEP-TH 9812136;%%
%``Tree amplitudes and two-loop counterterms in D = 11 supergravity,''
  Phys.\ Rev.\ D {\bf 62}, 084010 (2000)
  [arXiv:hep-th/0002241].
  %%CITATION = HEP-TH 0002241;%%

\item{[50]}
A.~Dasgupta, H.~Nicolai and J.~Plefka,
  %``Vertex operators for the supermembrane,''
  JHEP {\bf 0005}, 007 (2000)
  [arXiv:hep-th/0003280].
  %%CITATION = HEP-TH 0003280;%%

\item{[51]}
    A.~A.~Tseytlin,
% {\sl R**4 terms in 11 dimensions and conformal anomaly of (2,0) theory},
    Nucl.\ Phys.\ B {\bf 584}, 233 (2000)
    [arXiv:hep-th/0005072].
    %%CITATION = HEP-TH 0005072;%%

\item{[52]}
M.~Cederwall, U.~Gran, M.~Nielsen and B.~E.~W.~Nilsson,
  %``Manifestly supersymmetric M-theory,''
  JHEP {\bf 0010}, 041 (2000)
  [arXiv:hep-th/0007035].
  %%CITATION = HEP-TH 0007035;%%

\item{[53]} J. Plefka,
%{\sl Vertex operators for the supermembrane and background field
%matrix theory},
Int. J. Mod. Phys. {\bf A16} (2001) 660, {\tt
hep-th/0009193}

\item{[54]}
    K.~Peeters, P.~Vanhove and A.~Westerberg,
%  {\sl Supersymmetric higher-derivative actions in ten and eleven dimensions,
%  the associated superalgebras and their formulation in superspace},
    Class.\ Quant.\ Grav.\  {\bf 18} (2001) 843
    [arXiv:hep-th/0010167].
    %%CITATION = HEP-TH 0010167;%%

\item{[55]}
P.~S.~Howe and D.~Tsimpis,
  %``On higher-order corrections in M theory,''
  JHEP {\bf 0309}, 038 (2003)
  [arXiv:hep-th/0305129].
  %%CITATION = HEP-TH 0305129;%%

\item{[56]}
    P.~S.~Howe,
% {\sl R**4 terms in supergravity and M-theory,},
    arXiv:hep-th/0408177.
    %%CITATION = HEP-TH 0408177;%%

\item{[57]}
    G.~Policastro and D.~Tsimpis,
    %``R**4, purified,''
    arXiv:hep-th/0603165.
    %%CITATION = HEP-TH 0603165;%%

\item{[58]}
    T.~Damour and H.~Nicolai,
    %``Higher order M theory corrections and the Kac-Moody algebra E(10),''
    arXiv:hep-th/0504153.
    %%CITATION = HEP-TH 0504153;%%

\item{[59]}
T.~Damour, A.~Hanany, M.~Henneaux, A.~Kleinschmidt and H.~Nicolai,
    %``Curvature corrections and Kac-Moody compatibility conditions,''
    Gen.\ Rel.\ Grav.\  {\bf 38}, 1507 (2006)
    [arXiv:hep-th/0604143].
    %%CITATION = HEP-TH 0604143;%%

\item{[60]}  T.~Damour and M.~Henneaux,
    %``E(10), BE(10) and arithmetical chaos in superstring cosmology,''
    Phys.\ Rev.\ Lett.\  {\bf 86}, 4749 (2001)
    [arXiv:hep-th/0012172].
    %%CITATION = HEP-TH 0012172;%%

\item{[61]}  T.~Damour, M.~Henneaux and H.~Nicolai,
    %``E(10) and a 'small tension expansion' of M theory,''
    Phys.\ Rev.\ Lett.\  {\bf 89}, 221601 (2002)
    [arXiv:hep-th/0207267].
    %%CITATION = HEP-TH 0207267;%%

\item{[62]}
I.~Schnakenburg and P.~C.~West,
  %``Kac-Moody symmetries of IIB supergravity,''
  Phys.\ Lett.\ B {\bf 517}, 421 (2001)
  [arXiv:hep-th/0107181].
  %%CITATION = HEP-TH 0107181;%%

\item{[63]}
B.~Pioline and A.~Waldron,
%``Automorphic forms: A physicist's survey,''
arXiv:hep-th/0312068.
%%CITATION = HEP-TH 0312068;%%

\item{[64]}D.~Bump, {\it Automorphic forms in GL(3,R}),
Lecture notes in mathematics, Springer-verlag, 1984.

\item{[65]}  P.~C.~West,
    %``E(11) and M theory,''
    Class.\ Quant.\ Grav.\  {\bf 18}, 4443 (2001)
    [arXiv:hep-th/0104081].
    %%CITATION = HEP-TH 0104081;%%

\item{[66]} S.~R.~Coleman, J.~Wess and B.~Zumino,
    %``Structure Of Phenomenological Lagrangians. 1,''
    Phys.\ Rev.\  {\bf 177}, 2239 (1969).
    %%CITATION = PHRVA,177,2239;%%
    C.~G.~.~Callan, S.~R.~Coleman, J.~Wess and B.~Zumino,
    %``Structure Of Phenomenological Lagrangians. 2,''
    Phys.\ Rev.\  {\bf 177}, 2247 (1969).
    %%CITATION = PHRVA,177,2247;%%

\end